\documentclass[aps,prl,twocolumn, superscriptaddress,longbibliography,floatfix]{revtex4-2}

\usepackage[no-math]{fontspec}
\usepackage{graphicx,xcolor}
\usepackage{amsmath,amssymb,amsthm}
\usepackage[small,bf,justification=raggedright]{caption}
\usepackage{physics}
\usepackage[subrefformat=parens, labelformat=parens,singlelinecheck=false, justification=raggedright]{subcaption}
\usepackage[pdfencoding=auto]{hyperref}
\hypersetup{colorlinks=true,allcolors=black}
\usepackage[export]{adjustbox}
\usepackage{multirow}

\begin{document}

\title{Subdimensional linear-optical quantum computation: \\
from qudit resource states to qubit quantum computation}
% \date{\today}
\author{Tomohiro Yamazaki}
\email{tomohiro.yamazaki@ntt.com}
\affiliation{Basic Research Laboratories \& NTT Research Center for Theoretical Quantum Information, NTT Inc., Atsugi, Kanagawa 243-0198, Japan}

\begin{abstract}
In this paper, we propose a linear-optical quantum computation scheme that starts from high-dimensionally entangled qudits but performs quantum computation on qubits defined in their subspaces.
This approach enables us to increase the success probability of linear-optical fusions from $50\%$ to $1-1/d$ for qudits of dimension $d$ without relying on existing approaches using ancilla photons and quantum codes.
In particular, we show that a 2D cluster state of qubits can be generated from $5$-qudit 1D cluster states using \textit{pairwise fusion gates} while preserving their boosted success probability.
We numerically demonstrate that this approach can improve the loss tolerance of a percolation-based scheme while keeping the photon number per resource state constant.

\end{abstract}
\maketitle

\textit{Introduction---}
Photonic quantum computation remains a promising candidate for scalable quantum computation because of the inherently scalable nature of light~\cite{Kok_Rev.Mod.Phys.2007,Rudolph_APLPhotonics2017}.
However, since photons do not directly interact with each other, entangling gates need to be implemented using interference between photons or interactions with additional matter qubits.
The interference-based entangling gates, called linear-optical fusion gates, are inevitably probabilistic \cite{Lutkenhaus_Phys.Rev.A1999}, and their typical success probability is only 50\% \cite{Calsamiglia_Appl.Phys.B2001}.
This probabilistic nature necessitates many active switches and incurs enormous resource overhead \cite{Li_Phys.Rev.X2015,Gimeno-Segovia_NewJ.Phys.2017}, which makes it challenging to build photonic quantum computers.

Recent theoretical proposals, such as percolation-based schemes \cite{Kieling_Phys.Rev.Lett.2007,Gimeno-Segovia_Phys.Rev.Lett.2015,Pant_Nat.Commun.2019} and fusion-based schemes \cite{Bartolucci_Nat.Commun.2023}, assume fixed-size entangled states as resource states and combine them into a large entangled state (in effect in the latter case) using linear-optical fusions.
These resource states are normally generated by multiplexed linear-optical circuits~\cite{Bartolucci__2021} or quantum emitters~\cite{Lindner_Phys.Rev.Lett.2009,Schwartz_Science2016,Thomas_Nature2022}.
These architectures successfully reduce active routing and resource overhead, while a high success probability is required for each linear-optical fusion.
So far, such a high fusion probability has been achieved using two approaches: the use of ancilla photons \cite{Grice_Phys.Rev.A2011,Ewert_Phys.Rev.Lett.2014,Olivo_Phys.Rev.A2018} including squeezing~\cite{Zaidi_Phys.Rev.Lett.2013} and the use of quantum error-correcting codes \cite{Ewert_Phys.Rev.Lett.2016,Lee_Phys.Rev.A2019,Song_Phys.Rev.Lett.2024, Pankovich_Phys.Rev.Lett._2024}. In the former approach, the success probability can be increased to $75\%$ by supplying a Bell state \cite{Grice_Phys.Rev.A2011} or four single photons \cite{Ewert_Phys.Rev.Lett.2014} as additional inputs to the linear-optical circuit for each fusion.

In these architectures, photon loss and resource-state preparation are major experimental bottlenecks. To achieve a higher success probability, the ancilla-based approach needs more ancilla photons as inputs. This makes each fusion more vulnerable to photon loss because the linear-optical fusion fails when at least one input photon is lost.
On the other hand, the encoding-based approach can be made resilient to photon loss, in principle, up to the loss threshold of $50\%$ \cite{Varnava_Phys.Rev.Lett.2006, Varnava_NewJ.Phys.2007, Azuma_Nat.Commun.2015, Hilaire_PRXQuantum_2023}.
However, higher loss tolerance requires more photons to encode a single qubit.
Even a small increase in the number of physical qubits would lead to a significant resource overhead, as quantified in Ref.~\cite{Bartolucci_2025} in terms of the average number of 3-qubit GHZ states consumed.

Recently, a third approach based on a new type of entangling operation, called a pairwise fusion gate (PFG), has been proposed~\cite{Yamazaki_Phys.Rev.Lett._2025}.
The input of a PFG is a pair of qudits of dimension $d$, rather than qubits, and the PFG projects the input state onto a Bell state defined in a randomly chosen two-qubit subspace.
In this definition, the success probability is $1-1/d$ without ancilla photons, and it can be further enhanced with ancilla photons~\cite{Yamazaki_Phys.Rev.Lett._2025} and squeezing~\cite{Laha__2026}.
It immediately follows that a PFG can generate a qubit Bell state from a pair of qudit Bell states with this probability.
In contrast, it is nontrivial whether we can generate an entangled qubit state by applying multiple PFGs to qudit resource states because each PFG selects the effective qubit subspaces randomly.
In particular, while it was shown in~\cite{Yamazaki_Phys.Rev.Lett._2025} that the success probability of entanglement swapping can be repeatedly enhanced using PFGs, it remains open whether universal quantum computation is possible using PFGs while preserving the boosted success probability.

\begin{figure*}[t]
  \centering
  \includegraphics[width=\textwidth]{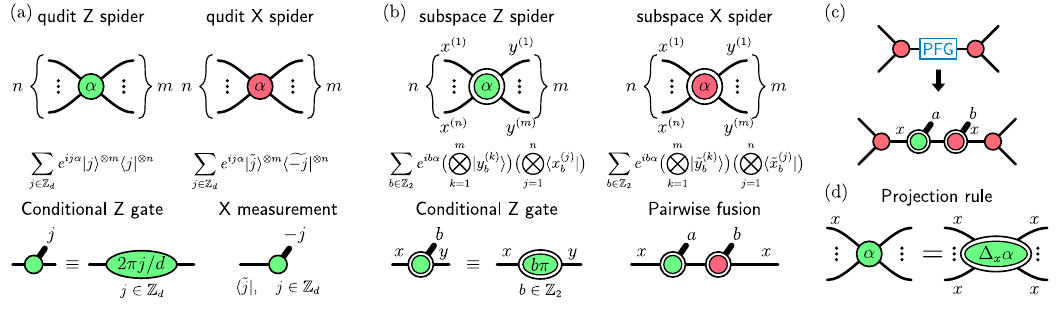}
  \caption{ZX-diagrammatic notations for (a) qudits and (b) two-dimensional subspaces, where $\ket{\tilde{j}} = \sqrt{d^{-1}}\sum_{k=0}^{d-1} e^{2\pi i jk/d}\ket{k}$ and $\ket{\tilde{x}_a}= \sqrt{2^{-1}}\sum_{b\in \mathbb{Z}_2}(-1)^{ab}\ket{x_b}$ for $x\in P$. (c) Diagrammatic representation of pairwise fusion of a pair of 3-qudit GHZ states. (d) Additional rewrite rule, called the projection rule.}
  \label{fig:definitions}
\end{figure*}

In this paper, we show that 2D cluster states of qubits can be generated from 5-qudit 1D cluster states using PFGs and single-qudit measurements while preserving the boosted success probability. 
Thus, universal quantum computation is possible by performing measurement-based quantum computation on the generated qubit state~\cite{Raussendorf_Phys.Rev.Lett._2001,Briegel_NaturePhys_2009}.
This result demonstrates that using qudit resource states for qubit quantum computation constitutes a third approach to increasing the fusion success probability in linear-optical quantum computation.
It is also important that the protocol starts from 1D cluster states because they can be generated deterministically with a single quantum emitter~\cite{Lindner_Phys.Rev.Lett.2009,Hilaire_Quantum2023}.
Although linear-optical generation of GHZ states \cite{Paesani_Phys.Rev.Lett.2021,Chin_Phys.Rev.Lett._2024} and their fusions \cite{Luo_Phys.Rev.Lett._2019,Zhang_Phys.Rev.A_2019,Hu_Phys.Rev.Lett._2020,Bharos_Quantum_2025,Ustun_Phys.Rev.Appl._2025} are still less efficient for qudits,
a multi-level quantum emitter should be able to generate high-dimensional entangled states directly~\cite{Raissi_PRXQuantum2024}.
Therefore, this ``subdimensional'' approach based on pairwise fusions may offer practical advantages especially in emitter-based schemes~\cite{Paesani_Phys.Rev.Lett._2023, Hilaire_Quantum2023, Lobl_Quantum_2024,Wein_PRXQuantum_2025,Chan_npjQuantumInf_2026}.

\textit{Pairwise fusion---}
We begin by reviewing the results in \cite{Yamazaki_Phys.Rev.Lett._2025}.
A family of linear-optical circuits using $2(2^k-1)$ ancilla photons for integers $k\geq 0$ has been shown to implement two-qudit measurements, called PFGs, on $d$-rail-encoded qudits.
For each $k$, the corresponding PFG is specified by Kraus operators
\begin{equation}\label{eq:definition_pfg}
    \{{\sqrt{d^{-k}}}\bra{i}\bra{i}, \bra*{\psi^+_{i,j}}, \bra*{\psi^-_{i,j}}, \sqrt{c_k}\bra*{\phi^+_{i,j}}, \sqrt{c_k}\bra*{\phi^-_{i,j}} \}
\end{equation}
for $i\in \mathbb{Z}_d=\{0,1,\cdots, d-1\}$ and $i<j \in \mathbb{Z}_d$ with $c_k=\sum_{p=1}^k d^{-p}$. Here, we call $\ket*{\psi^\pm_{i,j}}=\sqrt{2^{-1}}(\ket{i}\ket{j}\pm\ket{j}\ket{i})$ and $\ket*{\phi^\pm_{i,j}}=\sqrt{2^{-1}}(\ket{i}\ket{i}\pm\ket{j}\ket{j})$ pairwise Bell states.
In particular, no ancilla is used for $k=0$, and an ancillary high-dimensional Bell state $\ket{\phi} = \sqrt{d^{-1}}\sum_{i\in \mathbb{Z}_d} \ket{i}\ket{i}$ is used for $k=1$.
We define the success probability of PFGs as the probability of the maximally mixed input state being projected onto one of the pairwise Bell states.
For the PFG in Eq.~\eqref{eq:definition_pfg}, this probability is $p_k = 1-d^{-(k+1)}$.
Throughout this paper, we assume that the addition of elements in $\mathbb{Z}_d$ ($\mathbb{Z}_2$) is modulo $d$ ($2$).

This success probability applies when the pair of input qudits is maximally entangled with an external system.
For example, when we apply a PFG between a pair of high-dimensional Bell states $\ket{\phi}\ket{\phi}$, we obtain one of the pairwise Bell states with total success probability $p_k$.
The measurement outcome of a successful PFG can be specified by $x \in P = \{(i,j)\mid i \neq j \in \mathbb{Z}_d\}$ and $(a,b) \in \mathbb{Z}_2^2$ as a projection onto 
$\sqrt{2^{-1}}(\ket{x_0}\ket{x_b} +(-1)^a \ket{x_1}\ket{x_{1+b}})$.
Let us consider applying a PFG between a pair of states $\ket{C}\ket{C}$ for a 3-qudit entangled state $\ket{C} =d^{-1}\sum_{i,j \in \mathbb{Z}_d}\ket{i}\ket{-i-j}\ket{j}$~\footnote{Compared with \cite{Yamazaki_Phys.Rev.Lett._2025}, we replace $\ket{i+j}$ with $\ket{-i-j}$ in the definition of $\ket{C}$. This makes the state symmetric under permutations of the three qudits and allows it to be represented by a single 3-leg qudit X spider.}.
This state is locally equivalent to a 3-qudit GHZ state.
When the fusion succeeds, we obtain the state
\begin{align}\label{eq:Bell_state} 
    \ket*{\Phi^{a,b}_{x,x}}
    &= \frac{1}{d\sqrt{2}}\sum_{i,j \in \mathbb{Z}_d}\sum_{c \in \mathbb{Z}_2}(-1)^{ac}\notag \\
    &\quad\times\ket{i}\ket{-i-x_c}\ket{-x_{b+c}-j}\ket{j},
\end{align}
with total success probability $p_k$, where $x\in P$ and $(a,b)\in \mathbb{Z}_2^2$ represent the measurement outcome.

\textit{Diagrammatic representation---}
As Eq.~\eqref{eq:Bell_state} shows, the qudit states obtained by applying PFGs to graph states may be neither graph states nor stabilizer states, which makes it cumbersome to keep track of these state transformations.
However, we find that these states can still be represented concisely by using an extension of the ZX calculus, which is a diagrammatic language consisting of ZX diagrams and rules for rewriting them \cite{Coecke_NewJ.Phys._2011,vandeWetering_2020}.
For qudit systems, we define Z and X spiders as in Fig.~\ref{fig:definitions}(a)~\footnote{We define qudit X spiders using $\bra{\tilde{-j}}$ instead of $\bra{\tilde{j}}$ similar to the notation in~\cite{deBeaudrap_49thInt.Symp.Math.Found.Comput.Sci.MFCS2024_2024} to make diagrams symmetric under interchanges of kets and bras.
In addition, we let a phase of a qudit spider be a scalar, instead of a vector.}.
We also adopt the notation for quantum instruments used in~\cite{Bombin_Quantum2024}, as illustrated for X measurements in Fig.~\ref{fig:definitions}(a). 

We also introduce subspace Z and X spiders defined in Fig.~\ref{fig:definitions}(b). These Z and X spiders are the same as those for qubit systems except that the qubit for each leg labeled by $x \in P$ is defined in the subspace spanned by $\{\ket{x_0}, \ket{x_1}\}$.
A pairwise fusion can be represented using this notation as in Fig.~\ref{fig:definitions}(b) including its measurement outcome $x\in P$ and $(a,b) \in \mathbb{Z}_2^2$.
These subspace spiders can be connected to qudit spiders with labeled legs by considering each leg labeled by $x\in P$ as the projector $\dyad{x_0}+\dyad{x_1}$.
We omit labels on legs connecting subspace spiders because the processes they represent are independent of these labels.
Using this notation, the generation of $\ket*{\Phi^{a,b}_{x,x}}$ from $\ket{C}\ket{C}$ can be graphically represented as in Fig.~\ref{fig:definitions}(c), where each 3-leg spider corresponds to the state $\ket{C}$.

These diagrams can also be manipulated using rewrite rules, as in the standard ZX calculus.
An important property is that we can apply rewrite rules of the qudit (qubit) ZX calculus~\footnote{In our definition of qudit X spiders, some rewrite rules need to be modified as in~\cite{deBeaudrap_49thInt.Symp.Math.Found.Comput.Sci.MFCS2024_2024}. For example, a phase-free qudit X spider with one input and one output represents $\sum_{j\in \mathbb{Z}_d} \ket{j}\bra{-j}$, which differs from the identity for $d>2$. This would cause little confusion as we mainly use rewrite rules for subspace ZX diagrams.} to parts of diagrams consisting only of qudit (subspace) ZX diagrams.
% We assume some familiarity with qubit ZX calculus as we mostly apply rewrite rules only to subspace ZX diagrams.
We also have an immediate relation between different types of diagrams, called the projection rule in Fig.~\ref{fig:definitions}(d): Once one of the legs of a qudit Z spider is labeled by $x\in P$, it becomes a subspace Z spider with all legs labeled by $x$. In addition, a phase $\alpha$ reduces to $\Delta_x \alpha$ for $\Delta_x=x_1-x_0$, up to a global phase.

\textit{Adaptive single-qudit measurements---}
The treatment of qudit Z spiders with labeled legs is almost trivial because of the projection rule.
In contrast, qudit X spiders with labeled legs induce a structure beyond qubit and qudit graph states.
Let us consider a 3-leg X spider whose two legs are labeled by $x, y \in P$. Up to normalization, the corresponding 3-qudit state is
\begin{equation}\label{eq:labeled_Xspider}
    \sum_{a,b \in \mathbb{Z}_2}\ket{x_a}\ket{y_b}\ket{-(x_a +y_b)}.
\end{equation}
Importantly, the third qudit, corresponding to the unlabeled leg, is already restricted to the two-, three-, or four-dimensional subspace depending on whether (i) $x_0+y_0 = x_1 + y_1$ and/or (ii) $x_0+y_1 = x_1 + y_0$ hold. 
In particular, both conditions are satisfied simultaneously iff $\Delta_x = \Delta_y = d/2 \pmod{d}$.
For this reason, quantum operations on this third qudit need to be designed for all these cases.

\begin{figure}[t]
  \centering
  \includegraphics[]{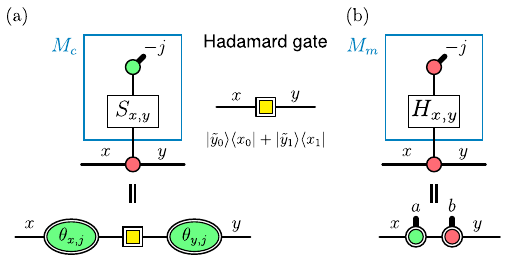}
  \caption{Definitions and corresponding diagrammatic transformations of (a) connecting measurement $M_c$ and (b) merging measurement $M_m$. Here, $\theta_{v,j}=\pi/2 + 2\pi j \Delta_v /d$ for $v=x,y$, and $j=-(x_a+y_{a+b})$ with $a=0$ for the case of $x_0+y_{0+b}=x_1+y_{1+b}$.}
  \label{fig:adaptive-measurements}
\end{figure}

Our aim here is to design a single-qudit measurement whose action in this setting is independent of the choice of indices $x,y \in P$.
In the 3-leg spider, a single-qudit measurement on the unlabeled leg acts as a two-qubit operation on the labeled legs through Eq.~\eqref{eq:labeled_Xspider}.
However, for instance, when the condition (ii) holds, we can only apply a two-qubit operation that acts on $\ket{x_0}\ket{y_1}$ and $\ket{x_1}\ket{y_0}$ identically by measuring the unlabeled leg. 
Despite this restriction, we show that adaptive single-qudit measurements, called connecting measurement $M_c$ and merging measurement $M_m$, produce the transformations shown in Fig.~\ref{fig:adaptive-measurements} regardless of whether conditions (i) and/or (ii) hold. As shown in Fig.~\ref{fig:adaptive-measurements}(a),
$M_c$ connects the two labeled legs with a subspace Hadamard gate inserted, similar to the Y measurement on graph states, with outcome-dependent phase factors.
On the other hand, as shown in Fig.~\ref{fig:adaptive-measurements}(b), $M_m$ directly connects the two labeled legs, similar to the X measurement on graph states, with outcome-dependent X and Z gates.
Once these transformations are diagrammatically represented, we can include these measurements in our set of diagrammatic rules.

The connecting measurement $M_c$ consists of a qudit X measurement following a single-qudit gate $S_{x,y}$ defined for $x,y \in P$ as $S_{x,y}\ket{-j} = i \ket{-j}$ when $j \in \{x_0+y_1, x_1 + y_0\} $ and otherwise $S_{x,y}\ket{-j} = \ket{-j}$.
Implementing $S_{x,y}$ on the unlabeled leg corresponds to the two-qubit operation of $\text{diag}(1,i,i,1)=(S\otimes S)\mathrm{CZ} $ on the labeled legs, regardless of whether conditions (i) and (ii) are satisfied.
The subsequent X measurement removes the unlabeled leg while introducing a phase factor $(e^{-2\pi i/d})^{-j(x_a+y_b)}$ for each term $\ket{x_a}\ket{y_b}$ when the measurement outcome is $j$.
As a result, we obtain the transformation in Fig.~\ref{fig:adaptive-measurements}(a).
On the other hand, the merging measurement $M_m$ is the Z measurement after applying $H_{x,y}$ defined for $x,y \in P$ as 
\begin{align}\label{eq:pairwise_Hadamard}
    &H_{x,y}\ket{-(x_a+y_{a+b})} = \notag \\
        &\quad \frac{1}{\sqrt{2}}(\ket{-(x_0+y_b)} + (-1)^a \ket{-(x_1 + y_{1+b})}) 
\end{align}
for $a,b \in \mathbb{Z}_2$ such that $x_0+y_b \neq x_1+y_{b+1}$ and otherwise $H_{x,y}\ket{-j} = \ket{-j}$.
In other words, for each $b\in \mathbb{Z}_2$, $H_{x,y}$ acts on the subspace spanned by $\{\ket{-(x_0+y_b)},\ket{-(x_1+y_{b+1})}\}$ as a Hadamard gate only when the subspace is two-dimensional.
From Eq.~\eqref{eq:labeled_Xspider},
measuring the unlabeled leg in the Z basis corresponds to projecting the labeled legs into a subspace spanned by $\{\ket{x_a}\ket{y_b} \mid j=-(x_a+y_b)\}$ for the measurement outcome $j$.
After applying $H_{x,y}$, obtaining $j=-(x_a+y_{a+b})$ in the Z measurement corresponds to the projection onto $\ket{x_0}\ket{y_b}+(-1)^a \ket{x_1}\ket{y_{b+1}}$ even when $x_0+y_b \neq x_1+y_{b+1}$.
This transformation is summarized in Fig.~\ref{fig:adaptive-measurements}(b).

\begin{figure*}[t]
  \centering
  \includegraphics[width=\textwidth]{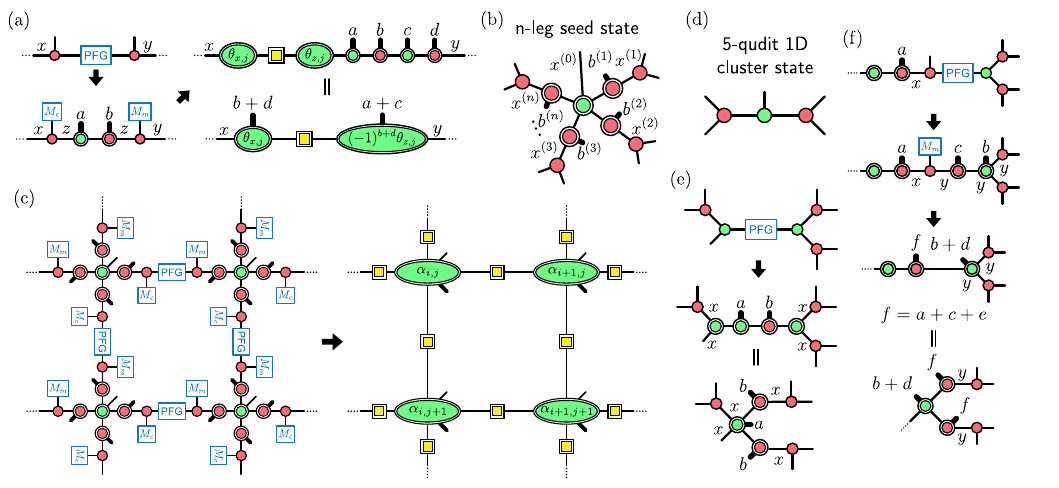}
  \caption{(a) Boosted fusion gate on a pair of X spiders. (b) Diagrammatic representation of an $n$-leg seed state.  (c) Generation of a 2D cluster state from 4-leg seed states. (d) Diagrammatic representation of a 5-qudit 1D cluster state. (e) Generation of a 3-leg seed state. (f) Extension from an $n$-leg seed state to an $(n+1)$-leg seed state.}
  \label{fig:star-and-lattice}
\end{figure*}

\textit{Graph state generation---}
Using the above results, we diagrammatically show that a 2D qubit cluster state can be generated from 5-qudit 1D cluster states.
Let us first consider the process shown in Fig.~\ref{fig:star-and-lattice}(a).
The input is a pair of 3-leg X spiders, each of which is connected to other diagrams with a leg labeled by $x, y \in P$ and has two unlabeled open legs.
This 3-leg X spider labeled by $x\in P$ can be regarded as an encoding of an effective qubit into a pair of entangled qudits as $\ket{x_a} \rightarrow \sqrt{d^{-1}}\sum_{i\in \mathbb{Z}_d}\ket{i}\ket{-(x_a+i)}$.
As in Fig.~\ref{fig:star-and-lattice}(a), we apply a PFG to a pair of qudits from the different spiders and then apply $M_m$ and $M_c$ on the remaining two qudits.
A diagrammatic calculation shows that a pair of effective qubits labeled by $x, y \in P$ is connected with a Hadamard gate inserted, up to the outcome-dependent phase factors.
This effective fusion gate on a pair of qubits succeeds with probability $p_k$.
Similarly, when we apply $M_m$ to both remaining qudits, a pair of effective qubits is connected without a Hadamard gate.
This process is already used in the boosted entanglement swapping proposed in~\cite{Yamazaki_Phys.Rev.Lett._2025}.

Then, we consider the $n$-leg seed states defined as in Fig.~\ref{fig:star-and-lattice}(b).
This state contains an effective qubit, represented by an open leg of a subspace Z spider. This spider is connected to $n$ qudit X spiders, each with two unlabeled open legs. A subspace X spider may optionally be inserted between the central Z spider and each qudit X spider.
This seed state has two important properties: We can apply the boosted fusion gate in Fig.~\ref{fig:star-and-lattice}(a) to each X spider to connect a pair of seed states, and the outcome-dependent phase factors can be canceled out by applying local unitary operations to the center effective qubits.
As shown in Fig.~\ref{fig:star-and-lattice}(c), let us consider putting 4-leg seed states on a square lattice and applying the boosted fusion gate on every pair of X spiders.
When all the fusions succeed, the obtained state corresponds to the 2D cluster state of qubits, up to local phase factors, whose values are known from the measurement outcomes.
More generally, we can generate arbitrary qubit graph states by preparing $n$-leg seed states with a sufficiently large $n$.

Next, we consider preparation of seed states from graph states of qudits.
Up to local unitary operations, 1D cluster states are graphically represented as alternating chains of X and Z spiders, with one open leg on each internal spider and two on each endpoint spider, as shown in Fig.~\ref{fig:star-and-lattice}(d) for five qudits.
As shown in Fig.~\ref{fig:star-and-lattice}(e), a 3-leg seed state is generated by applying a PFG between 5-qudit and 4-qudit 1D cluster states. 
To increase the number of legs in a seed state, we use the process described in Fig.~\ref{fig:star-and-lattice}(f) consisting of an application of a PFG followed by a merging measurement $M_m$.
This process generates an $(n+1)$-leg seed state from an $n$-leg seed state and a 5-qudit 1D cluster state.
Since a 4-qudit cluster is prepared from a 5-qudit one, $n$ 5-qudit 1D cluster states are sufficient to generate an $(n+1)$-leg seed state
for $n\geq 3$.
Alternatively, we can directly prepare a 4-leg seed state by replacing the 4-qudit cluster state in the process in Fig.~\ref{fig:star-and-lattice}(e) with a 6-qudit \textit{caterpillar state}~\cite{Lobl_Phys.Rev.A_2025}, which belongs to the class of states generated by a single quantum emitter.

\textit{Percolation-based scheme---}
A remaining problem is that the probability that all the fusions succeed decreases exponentially in the number of fusions.
Instead of the repeat-until-success approach~\cite{Browne_Phys.Rev.Lett.2005,Duan_Phys.Rev.Lett.2005}, we use a percolation-based approach: after all fusion gates are applied simultaneously, we extract a perfect 2D cluster state from the resulting 2D cluster state with missing edges.
In the standard protocol using rotated-basis fusion gates~\cite{Gimeno-Segovia_Phys.Rev.Lett.2015},
the failure of each fusion corresponds to the removal of an edge of the underlying graph.
Therefore, the percolation-based protocol works only when the fusion success probability exceeds the bond-percolation threshold, which is $50\%$ for square lattices.
In addition, the protocol can be robust against photon loss by detaching lost qubits with single-qubit measurements on neighboring qubits, which corresponds to the removal of vertices.
Then, the percolation threshold of the corresponding model represents the loss threshold of the protocol~\cite{Lobl_Phys.Rev.Res._2024}, except that loss of the photons representing the graph vertices is neglected for simplicity.

To adapt the boosted fusion gate in Fig.~\ref{fig:star-and-lattice}(a) for the percolation-based protocol, we need to define the operations on the remaining qudits in the event of failure or photon loss during the first PFG.
When the first PFG fails, it effectively acts on the input qudits as a Z-basis measurement.
As a result, the remaining qudits reduce to effective qubits, whose labels are $(-x_0-i,-x_1-i)$ and $(-y_0-i,-y_1-i)$, respectively, for a failure outcome $\bra{i}\bra{i}$.
Then, we can either (i) perform subspace X measurements on these effective qubits to disconnect them from the center qubits or (ii) apply a rotated-basis subspace fusion gate, similar to Ref.~\cite{Gimeno-Segovia_Phys.Rev.Lett.2015}.
When photon loss occurs during the first PFG, that is, when at least one input photon is lost, we apply the qudit X-basis measurements to the remaining qudits to detach the lost qudits from the center qubits.
For the pair of single-qudit measurements after a successful PFG, we conservatively remove both neighboring vertices when at least one photon is lost. 

\begin{figure}[tbp]
  \centering
  \includegraphics[width=0.45\textwidth]{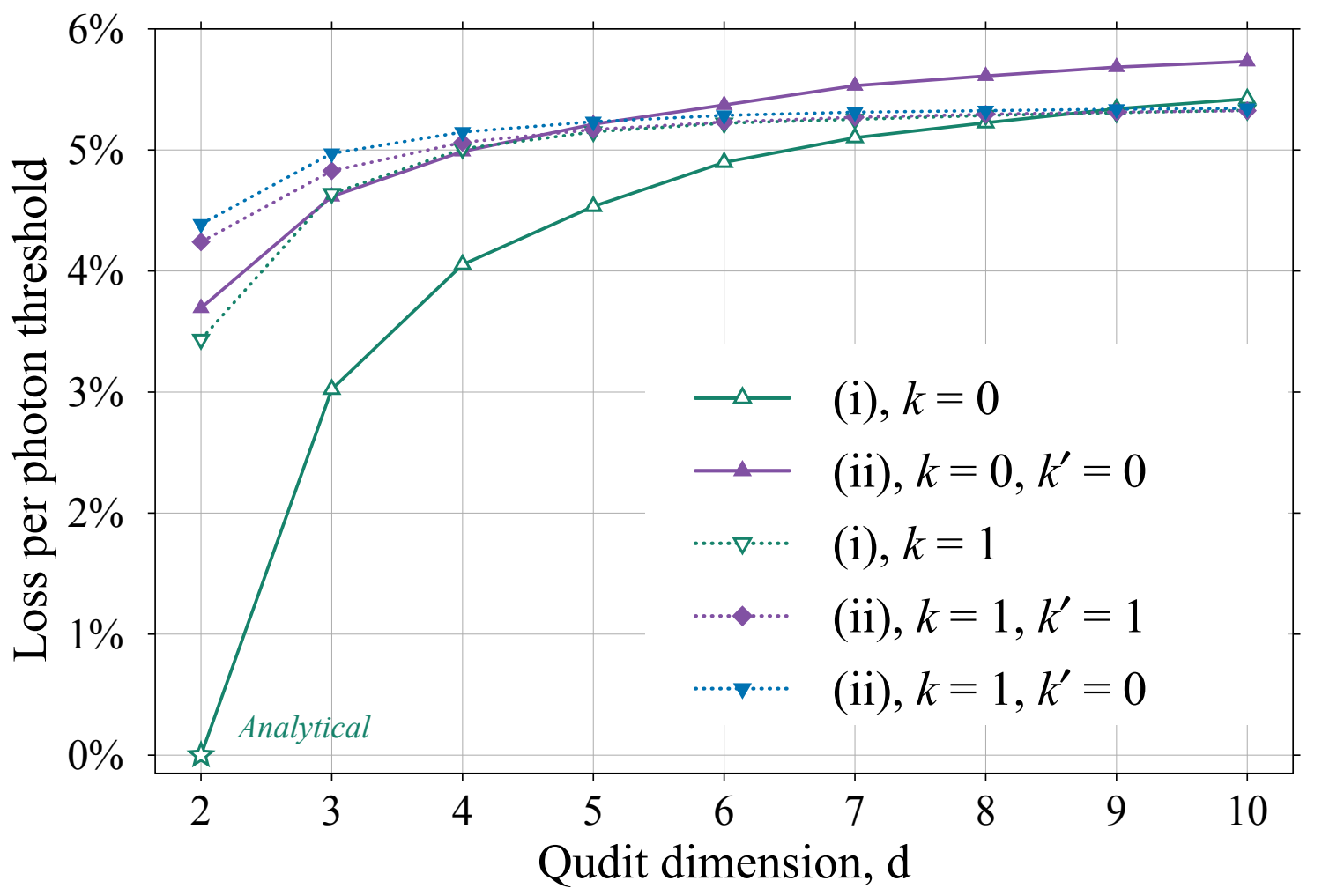}
  \caption{Loss tolerance of the percolation-based scheme on square lattices, where $k$ and $k'$ are parameters representing the use of ancilla photons in the PFG and subspace fusion gate, respectively. The numerical estimates are obtained from crossings of the percolation probabilities for different lattice sizes.}
  \label{fig:numerical_result}
\end{figure}

Figure~\ref{fig:numerical_result} shows the numerical results on the loss threshold of the above percolation-based scheme.
The benefit of the boosted success probability is reflected in an increased loss threshold, although the threshold seems to saturate for large $d$.
This is because our boosted fusion gates improve the fusion probability but not the loss probability as a pair of single-qudit measurements must succeed in addition to the first PFG.
The threshold saturates at a lower value when ancilla photons are used, because we assume the same loss probability for the ancilla photons.
These results motivate investigations of encoded fusion schemes built natively around PFGs.

\textit{Conclusions---}
In this paper, we have presented a linear-optical quantum computation scheme based on pairwise fusions of high-dimensionally entangled resource states.
We have introduced a ZX-diagrammatic representation to describe quantum processes involving PFGs and two types of useful adaptive single-qudit measurements.
Using these measurements and PFGs, we show that quantum states called seed states can be generated from 5-qudit 1D cluster states, and these seed states can be fused with the boosted success probability to create arbitrary qubit graph states. Finally, we have demonstrated the feasibility of universal quantum computation with PFGs under the percolation model and numerically evaluated its loss tolerance.

\textit{Acknowledgements---}
The author thanks Koji Azuma for fruitful discussions.
This work was supported by JST Moonshot R\&D Program Grant Number JPMJMS256E.

\textit{Disclosure of AI use---}
The central ideas of this work were initially conceived by the author without AI assistance, several months before the author found that GPT-5.6 Pro could independently derive a similar graph-state generation protocol.
The numerical simulations were performed with Julia code generated by GPT-6 Astra, which was also used to assist with refining the mathematical arguments, preparing figures, and editing the manuscript.
The author has verified all scientific results and takes full
responsibility for them.

\bibliography{sdqc.bib}

\begin{thebibliography}{55}%
\makeatletter
\providecommand \@ifxundefined [1]{%
 \@ifx{#1\undefined}
}%
\providecommand \@ifnum [1]{%
 \ifnum #1\expandafter \@firstoftwo
 \else \expandafter \@secondoftwo
 \fi
}%
\providecommand \@ifx [1]{%
 \ifx #1\expandafter \@firstoftwo
 \else \expandafter \@secondoftwo
 \fi
}%
\providecommand \natexlab [1]{#1}%
\providecommand \enquote  [1]{``#1''}%
\providecommand \bibnamefont  [1]{#1}%
\providecommand \bibfnamefont [1]{#1}%
\providecommand \citenamefont [1]{#1}%
\providecommand \href@noop [0]{\@secondoftwo}%
\providecommand \href [0]{\begingroup \@sanitize@url \@href}%
\providecommand \@href[1]{\@@startlink{#1}\@@href}%
\providecommand \@@href[1]{\endgroup#1\@@endlink}%
\providecommand \@sanitize@url [0]{\catcode `\\12\catcode `\$12\catcode
  `\&12\catcode `\#12\catcode `\^12\catcode `\_12\catcode `\%12\relax}%
\providecommand \@@startlink[1]{}%
\providecommand \@@endlink[0]{}%
\providecommand \url  [0]{\begingroup\@sanitize@url \@url }%
\providecommand \@url [1]{\endgroup\@href {#1}{\urlprefix }}%
\providecommand \urlprefix  [0]{URL }%
\providecommand \Eprint [0]{\href }%
\providecommand \doibase [0]{https://doi.org/}%
\providecommand \selectlanguage [0]{\@gobble}%
\providecommand \bibinfo  [0]{\@secondoftwo}%
\providecommand \bibfield  [0]{\@secondoftwo}%
\providecommand \translation [1]{[#1]}%
\providecommand \BibitemOpen [0]{}%
\providecommand \bibitemStop [0]{}%
\providecommand \bibitemNoStop [0]{.\EOS\space}%
\providecommand \EOS [0]{\spacefactor3000\relax}%
\providecommand \BibitemShut  [1]{\csname bibitem#1\endcsname}%
\let\auto@bib@innerbib\@empty
%</preamble>
\bibitem [{\citenamefont {Kok}\ \emph {et~al.}(2007)\citenamefont {Kok},
  \citenamefont {Munro}, \citenamefont {Nemoto}, \citenamefont {Ralph},
  \citenamefont {Dowling},\ and\ \citenamefont
  {Milburn}}]{Kok_Rev.Mod.Phys.2007}%
  \BibitemOpen
  \bibfield  {author} {\bibinfo {author} {\bibfnamefont {P.}~\bibnamefont
  {Kok}}, \bibinfo {author} {\bibfnamefont {W.~J.}\ \bibnamefont {Munro}},
  \bibinfo {author} {\bibfnamefont {K.}~\bibnamefont {Nemoto}}, \bibinfo
  {author} {\bibfnamefont {T.~C.}\ \bibnamefont {Ralph}}, \bibinfo {author}
  {\bibfnamefont {J.~P.}\ \bibnamefont {Dowling}},\ and\ \bibinfo {author}
  {\bibfnamefont {G.~J.}\ \bibnamefont {Milburn}},\ }\bibfield  {title}
  {\bibinfo {title} {Linear optical quantum computing with photonic qubits},\
  }\href {https://doi.org/10.1103/RevModPhys.79.135} {\bibfield  {journal}
  {\bibinfo  {journal} {Reviews of Modern Physics}\ }\textbf {\bibinfo {volume}
  {79}},\ \bibinfo {pages} {135} (\bibinfo {year} {2007})}\BibitemShut
  {NoStop}%
\bibitem [{\citenamefont {Rudolph}(2017)}]{Rudolph_APLPhotonics2017}%
  \BibitemOpen
  \bibfield  {author} {\bibinfo {author} {\bibfnamefont {T.}~\bibnamefont
  {Rudolph}},\ }\bibfield  {title} {\bibinfo {title} {Why {{I}} am optimistic
  about the silicon-photonic route to quantum computing},\ }\href
  {https://doi.org/10.1063/1.4976737} {\bibfield  {journal} {\bibinfo
  {journal} {APL Photonics}\ }\textbf {\bibinfo {volume} {2}},\ \bibinfo
  {pages} {030901} (\bibinfo {year} {2017})}\BibitemShut {NoStop}%
\bibitem [{\citenamefont {L{\"u}tkenhaus}\ \emph {et~al.}(1999)\citenamefont
  {L{\"u}tkenhaus}, \citenamefont {Calsamiglia},\ and\ \citenamefont
  {Suominen}}]{Lutkenhaus_Phys.Rev.A1999}%
  \BibitemOpen
  \bibfield  {author} {\bibinfo {author} {\bibfnamefont {N.}~\bibnamefont
  {L{\"u}tkenhaus}}, \bibinfo {author} {\bibfnamefont {J.}~\bibnamefont
  {Calsamiglia}},\ and\ \bibinfo {author} {\bibfnamefont {K.-A.}\ \bibnamefont
  {Suominen}},\ }\bibfield  {title} {\bibinfo {title} {Bell measurements for
  teleportation},\ }\href {https://doi.org/10.1103/PhysRevA.59.3295} {\bibfield
   {journal} {\bibinfo  {journal} {Physical Review A}\ }\textbf {\bibinfo
  {volume} {59}},\ \bibinfo {pages} {3295} (\bibinfo {year}
  {1999})}\BibitemShut {NoStop}%
\bibitem [{\citenamefont {Calsamiglia}\ and\ \citenamefont
  {L{\"u}tkenhaus}(2001)}]{Calsamiglia_Appl.Phys.B2001}%
  \BibitemOpen
  \bibfield  {author} {\bibinfo {author} {\bibfnamefont {J.}~\bibnamefont
  {Calsamiglia}}\ and\ \bibinfo {author} {\bibfnamefont {N.}~\bibnamefont
  {L{\"u}tkenhaus}},\ }\bibfield  {title} {\bibinfo {title} {Maximum efficiency
  of a linear-optical {{Bell}}-state analyzer},\ }\href
  {https://doi.org/10.1007/s003400000484} {\bibfield  {journal} {\bibinfo
  {journal} {Applied Physics B}\ }\textbf {\bibinfo {volume} {72}},\ \bibinfo
  {pages} {67} (\bibinfo {year} {2001})}\BibitemShut {NoStop}%
\bibitem [{\citenamefont {Li}\ \emph {et~al.}(2015)\citenamefont {Li},
  \citenamefont {Humphreys}, \citenamefont {Mendoza},\ and\ \citenamefont
  {Benjamin}}]{Li_Phys.Rev.X2015}%
  \BibitemOpen
  \bibfield  {author} {\bibinfo {author} {\bibfnamefont {Y.}~\bibnamefont
  {Li}}, \bibinfo {author} {\bibfnamefont {P.~C.}\ \bibnamefont {Humphreys}},
  \bibinfo {author} {\bibfnamefont {G.~J.}\ \bibnamefont {Mendoza}},\ and\
  \bibinfo {author} {\bibfnamefont {S.~C.}\ \bibnamefont {Benjamin}},\
  }\bibfield  {title} {\bibinfo {title} {Resource {{Costs}} for
  {{Fault-Tolerant Linear Optical Quantum Computing}}},\ }\href
  {https://doi.org/10.1103/PhysRevX.5.041007} {\bibfield  {journal} {\bibinfo
  {journal} {Physical Review X}\ }\textbf {\bibinfo {volume} {5}},\ \bibinfo
  {pages} {041007} (\bibinfo {year} {2015})}\BibitemShut {NoStop}%
\bibitem [{\citenamefont {{Gimeno-Segovia}}\ \emph {et~al.}(2017)\citenamefont
  {{Gimeno-Segovia}}, \citenamefont {Cable}, \citenamefont {Mendoza},
  \citenamefont {Shadbolt}, \citenamefont {Silverstone}, \citenamefont
  {Carolan}, \citenamefont {Thompson}, \citenamefont {O'Brien},\ and\
  \citenamefont {Rudolph}}]{Gimeno-Segovia_NewJ.Phys.2017}%
  \BibitemOpen
  \bibfield  {author} {\bibinfo {author} {\bibfnamefont {M.}~\bibnamefont
  {{Gimeno-Segovia}}}, \bibinfo {author} {\bibfnamefont {H.}~\bibnamefont
  {Cable}}, \bibinfo {author} {\bibfnamefont {G.~J.}\ \bibnamefont {Mendoza}},
  \bibinfo {author} {\bibfnamefont {P.}~\bibnamefont {Shadbolt}}, \bibinfo
  {author} {\bibfnamefont {J.~W.}\ \bibnamefont {Silverstone}}, \bibinfo
  {author} {\bibfnamefont {J.}~\bibnamefont {Carolan}}, \bibinfo {author}
  {\bibfnamefont {M.~G.}\ \bibnamefont {Thompson}}, \bibinfo {author}
  {\bibfnamefont {J.~L.}\ \bibnamefont {O'Brien}},\ and\ \bibinfo {author}
  {\bibfnamefont {T.}~\bibnamefont {Rudolph}},\ }\bibfield  {title} {\bibinfo
  {title} {Relative multiplexing for minimising switching in linear-optical
  quantum computing},\ }\href {https://doi.org/10.1088/1367-2630/aa7095}
  {\bibfield  {journal} {\bibinfo  {journal} {New Journal of Physics}\ }\textbf
  {\bibinfo {volume} {19}},\ \bibinfo {pages} {063013} (\bibinfo {year}
  {2017})}\BibitemShut {NoStop}%
\bibitem [{\citenamefont {Kieling}\ \emph {et~al.}(2007)\citenamefont
  {Kieling}, \citenamefont {Rudolph},\ and\ \citenamefont
  {Eisert}}]{Kieling_Phys.Rev.Lett.2007}%
  \BibitemOpen
  \bibfield  {author} {\bibinfo {author} {\bibfnamefont {K.}~\bibnamefont
  {Kieling}}, \bibinfo {author} {\bibfnamefont {T.}~\bibnamefont {Rudolph}},\
  and\ \bibinfo {author} {\bibfnamefont {J.}~\bibnamefont {Eisert}},\
  }\bibfield  {title} {\bibinfo {title} {Percolation, {{Renormalization}}, and
  {{Quantum Computing}} with {{Nondeterministic Gates}}},\ }\href
  {https://doi.org/10.1103/PhysRevLett.99.130501} {\bibfield  {journal}
  {\bibinfo  {journal} {Physical Review Letters}\ }\textbf {\bibinfo {volume}
  {99}},\ \bibinfo {pages} {130501} (\bibinfo {year} {2007})}\BibitemShut
  {NoStop}%
\bibitem [{\citenamefont {{Gimeno-Segovia}}\ \emph {et~al.}(2015)\citenamefont
  {{Gimeno-Segovia}}, \citenamefont {Shadbolt}, \citenamefont {Browne},\ and\
  \citenamefont {Rudolph}}]{Gimeno-Segovia_Phys.Rev.Lett.2015}%
  \BibitemOpen
  \bibfield  {author} {\bibinfo {author} {\bibfnamefont {M.}~\bibnamefont
  {{Gimeno-Segovia}}}, \bibinfo {author} {\bibfnamefont {P.}~\bibnamefont
  {Shadbolt}}, \bibinfo {author} {\bibfnamefont {D.~E.}\ \bibnamefont
  {Browne}},\ and\ \bibinfo {author} {\bibfnamefont {T.}~\bibnamefont
  {Rudolph}},\ }\bibfield  {title} {\bibinfo {title} {From {{Three-Photon
  Greenberger-Horne-Zeilinger States}} to {{Ballistic Universal Quantum
  Computation}}},\ }\href {https://doi.org/10.1103/PhysRevLett.115.020502}
  {\bibfield  {journal} {\bibinfo  {journal} {Physical Review Letters}\
  }\textbf {\bibinfo {volume} {115}},\ \bibinfo {pages} {020502} (\bibinfo
  {year} {2015})}\BibitemShut {NoStop}%
\bibitem [{\citenamefont {Pant}\ \emph {et~al.}(2019)\citenamefont {Pant},
  \citenamefont {Towsley}, \citenamefont {Englund},\ and\ \citenamefont
  {Guha}}]{Pant_Nat.Commun.2019}%
  \BibitemOpen
  \bibfield  {author} {\bibinfo {author} {\bibfnamefont {M.}~\bibnamefont
  {Pant}}, \bibinfo {author} {\bibfnamefont {D.}~\bibnamefont {Towsley}},
  \bibinfo {author} {\bibfnamefont {D.}~\bibnamefont {Englund}},\ and\ \bibinfo
  {author} {\bibfnamefont {S.}~\bibnamefont {Guha}},\ }\bibfield  {title}
  {\bibinfo {title} {Percolation thresholds for photonic quantum computing},\
  }\href {https://doi.org/10.1038/s41467-019-08948-x} {\bibfield  {journal}
  {\bibinfo  {journal} {Nature Communications}\ }\textbf {\bibinfo {volume}
  {10}},\ \bibinfo {pages} {1070} (\bibinfo {year} {2019})}\BibitemShut
  {NoStop}%
\bibitem [{\citenamefont {Bartolucci}\ \emph {et~al.}(2023)\citenamefont
  {Bartolucci}, \citenamefont {Birchall}, \citenamefont {Bomb{\'i}n},
  \citenamefont {Cable}, \citenamefont {Dawson}, \citenamefont
  {{Gimeno-Segovia}}, \citenamefont {Johnston}, \citenamefont {Kieling},
  \citenamefont {Nickerson}, \citenamefont {Pant}, \citenamefont {Pastawski},
  \citenamefont {Rudolph},\ and\ \citenamefont
  {Sparrow}}]{Bartolucci_Nat.Commun.2023}%
  \BibitemOpen
  \bibfield  {author} {\bibinfo {author} {\bibfnamefont {S.}~\bibnamefont
  {Bartolucci}}, \bibinfo {author} {\bibfnamefont {P.}~\bibnamefont
  {Birchall}}, \bibinfo {author} {\bibfnamefont {H.}~\bibnamefont
  {Bomb{\'i}n}}, \bibinfo {author} {\bibfnamefont {H.}~\bibnamefont {Cable}},
  \bibinfo {author} {\bibfnamefont {C.}~\bibnamefont {Dawson}}, \bibinfo
  {author} {\bibfnamefont {M.}~\bibnamefont {{Gimeno-Segovia}}}, \bibinfo
  {author} {\bibfnamefont {E.}~\bibnamefont {Johnston}}, \bibinfo {author}
  {\bibfnamefont {K.}~\bibnamefont {Kieling}}, \bibinfo {author} {\bibfnamefont
  {N.}~\bibnamefont {Nickerson}}, \bibinfo {author} {\bibfnamefont
  {M.}~\bibnamefont {Pant}}, \bibinfo {author} {\bibfnamefont {F.}~\bibnamefont
  {Pastawski}}, \bibinfo {author} {\bibfnamefont {T.}~\bibnamefont {Rudolph}},\
  and\ \bibinfo {author} {\bibfnamefont {C.}~\bibnamefont {Sparrow}},\
  }\bibfield  {title} {\bibinfo {title} {Fusion-based quantum computation},\
  }\href {https://doi.org/10.1038/s41467-023-36493-1} {\bibfield  {journal}
  {\bibinfo  {journal} {Nature Communications}\ }\textbf {\bibinfo {volume}
  {14}},\ \bibinfo {pages} {912} (\bibinfo {year} {2023})}\BibitemShut
  {NoStop}%
\bibitem [{\citenamefont {Bartolucci}\ \emph {et~al.}(2021)\citenamefont
  {Bartolucci}, \citenamefont {Birchall}, \citenamefont {{Gimeno-Segovia}},
  \citenamefont {Johnston}, \citenamefont {Kieling}, \citenamefont {Pant},
  \citenamefont {Rudolph}, \citenamefont {Smith}, \citenamefont {Sparrow},\
  and\ \citenamefont {Vidrighin}}]{Bartolucci__2021}%
  \BibitemOpen
  \bibfield  {author} {\bibinfo {author} {\bibfnamefont {S.}~\bibnamefont
  {Bartolucci}}, \bibinfo {author} {\bibfnamefont {P.~M.}\ \bibnamefont
  {Birchall}}, \bibinfo {author} {\bibfnamefont {M.}~\bibnamefont
  {{Gimeno-Segovia}}}, \bibinfo {author} {\bibfnamefont {E.}~\bibnamefont
  {Johnston}}, \bibinfo {author} {\bibfnamefont {K.}~\bibnamefont {Kieling}},
  \bibinfo {author} {\bibfnamefont {M.}~\bibnamefont {Pant}}, \bibinfo {author}
  {\bibfnamefont {T.}~\bibnamefont {Rudolph}}, \bibinfo {author} {\bibfnamefont
  {J.}~\bibnamefont {Smith}}, \bibinfo {author} {\bibfnamefont
  {C.}~\bibnamefont {Sparrow}},\ and\ \bibinfo {author} {\bibfnamefont {M.~D.}\
  \bibnamefont {Vidrighin}},\ }\href
  {https://doi.org/10.48550/arXiv.2106.13825} {\bibinfo {title} {Creation of
  {{Entangled Photonic States Using Linear Optics}}}} (\bibinfo {year}
  {2021}),\ \Eprint {https://arxiv.org/abs/2106.13825} {arXiv:2106.13825
  [quant-ph]} \BibitemShut {NoStop}%
\bibitem [{\citenamefont {Lindner}\ and\ \citenamefont
  {Rudolph}(2009)}]{Lindner_Phys.Rev.Lett.2009}%
  \BibitemOpen
  \bibfield  {author} {\bibinfo {author} {\bibfnamefont {N.~H.}\ \bibnamefont
  {Lindner}}\ and\ \bibinfo {author} {\bibfnamefont {T.}~\bibnamefont
  {Rudolph}},\ }\bibfield  {title} {\bibinfo {title} {Proposal for pulsed
  on-demand sources of photonic cluster state strings},\ }\href
  {https://doi.org/10.1103/PhysRevLett.103.113602} {\bibfield  {journal}
  {\bibinfo  {journal} {Physical Review Letters}\ }\textbf {\bibinfo {volume}
  {103}},\ \bibinfo {pages} {113602} (\bibinfo {year} {2009})}\BibitemShut
  {NoStop}%
\bibitem [{\citenamefont {Schwartz}\ \emph {et~al.}(2016)\citenamefont
  {Schwartz}, \citenamefont {Cogan}, \citenamefont {Schmidgall}, \citenamefont
  {Don}, \citenamefont {Gantz}, \citenamefont {Kenneth}, \citenamefont
  {Lindner},\ and\ \citenamefont {Gershoni}}]{Schwartz_Science2016}%
  \BibitemOpen
  \bibfield  {author} {\bibinfo {author} {\bibfnamefont {I.}~\bibnamefont
  {Schwartz}}, \bibinfo {author} {\bibfnamefont {D.}~\bibnamefont {Cogan}},
  \bibinfo {author} {\bibfnamefont {E.~R.}\ \bibnamefont {Schmidgall}},
  \bibinfo {author} {\bibfnamefont {Y.}~\bibnamefont {Don}}, \bibinfo {author}
  {\bibfnamefont {L.}~\bibnamefont {Gantz}}, \bibinfo {author} {\bibfnamefont
  {O.}~\bibnamefont {Kenneth}}, \bibinfo {author} {\bibfnamefont {N.~H.}\
  \bibnamefont {Lindner}},\ and\ \bibinfo {author} {\bibfnamefont
  {D.}~\bibnamefont {Gershoni}},\ }\bibfield  {title} {\bibinfo {title}
  {Deterministic generation of a cluster state of entangled photons},\ }\href
  {https://doi.org/10.1126/science.aah4758} {\bibfield  {journal} {\bibinfo
  {journal} {Science}\ }\textbf {\bibinfo {volume} {354}},\ \bibinfo {pages}
  {434} (\bibinfo {year} {2016})}\BibitemShut {NoStop}%
\bibitem [{\citenamefont {Thomas}\ \emph {et~al.}(2022)\citenamefont {Thomas},
  \citenamefont {Ruscio}, \citenamefont {Morin},\ and\ \citenamefont
  {Rempe}}]{Thomas_Nature2022}%
  \BibitemOpen
  \bibfield  {author} {\bibinfo {author} {\bibfnamefont {P.}~\bibnamefont
  {Thomas}}, \bibinfo {author} {\bibfnamefont {L.}~\bibnamefont {Ruscio}},
  \bibinfo {author} {\bibfnamefont {O.}~\bibnamefont {Morin}},\ and\ \bibinfo
  {author} {\bibfnamefont {G.}~\bibnamefont {Rempe}},\ }\bibfield  {title}
  {\bibinfo {title} {Efficient generation of entangled multiphoton graph states
  from a single atom},\ }\href {https://doi.org/10.1038/s41586-022-04987-5}
  {\bibfield  {journal} {\bibinfo  {journal} {Nature}\ }\textbf {\bibinfo
  {volume} {608}},\ \bibinfo {pages} {677} (\bibinfo {year}
  {2022})}\BibitemShut {NoStop}%
\bibitem [{\citenamefont {Grice}(2011)}]{Grice_Phys.Rev.A2011}%
  \BibitemOpen
  \bibfield  {author} {\bibinfo {author} {\bibfnamefont {W.~P.}\ \bibnamefont
  {Grice}},\ }\bibfield  {title} {\bibinfo {title} {Arbitrarily complete
  {{Bell}}-state measurement using only linear optical elements},\ }\href
  {https://doi.org/10.1103/PhysRevA.84.042331} {\bibfield  {journal} {\bibinfo
  {journal} {Physical Review A}\ }\textbf {\bibinfo {volume} {84}},\ \bibinfo
  {pages} {042331} (\bibinfo {year} {2011})}\BibitemShut {NoStop}%
\bibitem [{\citenamefont {Ewert}\ and\ \citenamefont {{van
  Loock}}(2014)}]{Ewert_Phys.Rev.Lett.2014}%
  \BibitemOpen
  \bibfield  {author} {\bibinfo {author} {\bibfnamefont {F.}~\bibnamefont
  {Ewert}}\ and\ \bibinfo {author} {\bibfnamefont {P.}~\bibnamefont {{van
  Loock}}},\ }\bibfield  {title} {\bibinfo {title} {3/4-{{Efficient Bell}}
  measurement with passive linear optics and unentangled ancillae},\ }\href
  {https://doi.org/10.1103/PhysRevLett.113.140403} {\bibfield  {journal}
  {\bibinfo  {journal} {Physical Review Letters}\ }\textbf {\bibinfo {volume}
  {113}},\ \bibinfo {pages} {140403} (\bibinfo {year} {2014})}\BibitemShut
  {NoStop}%
\bibitem [{\citenamefont {Olivo}\ and\ \citenamefont
  {Grosshans}(2018)}]{Olivo_Phys.Rev.A2018}%
  \BibitemOpen
  \bibfield  {author} {\bibinfo {author} {\bibfnamefont {A.}~\bibnamefont
  {Olivo}}\ and\ \bibinfo {author} {\bibfnamefont {F.}~\bibnamefont
  {Grosshans}},\ }\bibfield  {title} {\bibinfo {title} {Ancilla-assisted linear
  optical {{Bell}} measurements and their optimality},\ }\href
  {https://doi.org/10.1103/PhysRevA.98.042323} {\bibfield  {journal} {\bibinfo
  {journal} {Physical Review A}\ }\textbf {\bibinfo {volume} {98}},\ \bibinfo
  {pages} {042323} (\bibinfo {year} {2018})}\BibitemShut {NoStop}%
\bibitem [{\citenamefont {Zaidi}\ and\ \citenamefont
  {Van~Loock}(2013)}]{Zaidi_Phys.Rev.Lett.2013}%
  \BibitemOpen
  \bibfield  {author} {\bibinfo {author} {\bibfnamefont {H.~A.}\ \bibnamefont
  {Zaidi}}\ and\ \bibinfo {author} {\bibfnamefont {P.}~\bibnamefont
  {Van~Loock}},\ }\bibfield  {title} {\bibinfo {title} {Beating the {{One-Half
  Limit}} of {{Ancilla-Free Linear Optics Bell Measurements}}},\ }\href
  {https://doi.org/10.1103/PhysRevLett.110.260501} {\bibfield  {journal}
  {\bibinfo  {journal} {Physical Review Letters}\ }\textbf {\bibinfo {volume}
  {110}},\ \bibinfo {pages} {260501} (\bibinfo {year} {2013})}\BibitemShut
  {NoStop}%
\bibitem [{\citenamefont {Ewert}\ \emph {et~al.}(2016)\citenamefont {Ewert},
  \citenamefont {Bergmann},\ and\ \citenamefont {{van
  Loock}}}]{Ewert_Phys.Rev.Lett.2016}%
  \BibitemOpen
  \bibfield  {author} {\bibinfo {author} {\bibfnamefont {F.}~\bibnamefont
  {Ewert}}, \bibinfo {author} {\bibfnamefont {M.}~\bibnamefont {Bergmann}},\
  and\ \bibinfo {author} {\bibfnamefont {P.}~\bibnamefont {{van Loock}}},\
  }\bibfield  {title} {\bibinfo {title} {Ultrafast {{Long-Distance Quantum
  Communication}} with {{Static Linear Optics}}},\ }\href
  {https://doi.org/10.1103/PhysRevLett.117.210501} {\bibfield  {journal}
  {\bibinfo  {journal} {Physical Review Letters}\ }\textbf {\bibinfo {volume}
  {117}},\ \bibinfo {pages} {210501} (\bibinfo {year} {2016})},\ \Eprint
  {https://arxiv.org/abs/1503.06777} {arXiv:1503.06777 [quant-ph]} \BibitemShut
  {NoStop}%
\bibitem [{\citenamefont {Lee}\ \emph {et~al.}(2019)\citenamefont {Lee},
  \citenamefont {Ralph},\ and\ \citenamefont {Jeong}}]{Lee_Phys.Rev.A2019}%
  \BibitemOpen
  \bibfield  {author} {\bibinfo {author} {\bibfnamefont {S.-W.}\ \bibnamefont
  {Lee}}, \bibinfo {author} {\bibfnamefont {T.~C.}\ \bibnamefont {Ralph}},\
  and\ \bibinfo {author} {\bibfnamefont {H.}~\bibnamefont {Jeong}},\ }\bibfield
   {title} {\bibinfo {title} {Fundamental building block for all-optical
  scalable quantum networks},\ }\href
  {https://doi.org/10.1103/PhysRevA.100.052303} {\bibfield  {journal} {\bibinfo
   {journal} {Physical Review A}\ }\textbf {\bibinfo {volume} {100}},\ \bibinfo
  {pages} {052303} (\bibinfo {year} {2019})}\BibitemShut {NoStop}%
\bibitem [{\citenamefont {Song}\ \emph {et~al.}(2024)\citenamefont {Song},
  \citenamefont {Kang}, \citenamefont {Kim},\ and\ \citenamefont
  {Lee}}]{Song_Phys.Rev.Lett.2024}%
  \BibitemOpen
  \bibfield  {author} {\bibinfo {author} {\bibfnamefont {W.}~\bibnamefont
  {Song}}, \bibinfo {author} {\bibfnamefont {N.}~\bibnamefont {Kang}}, \bibinfo
  {author} {\bibfnamefont {Y.-S.}\ \bibnamefont {Kim}},\ and\ \bibinfo {author}
  {\bibfnamefont {S.-W.}\ \bibnamefont {Lee}},\ }\bibfield  {title} {\bibinfo
  {title} {Encoded-{{Fusion-Based Quantum Computation}} for {{High Thresholds}}
  with {{Linear Optics}}},\ }\href
  {https://doi.org/10.1103/PhysRevLett.133.050605} {\bibfield  {journal}
  {\bibinfo  {journal} {Physical Review Letters}\ }\textbf {\bibinfo {volume}
  {133}},\ \bibinfo {pages} {050605} (\bibinfo {year} {2024})}\BibitemShut
  {NoStop}%
\bibitem [{\citenamefont {Pankovich}\ \emph {et~al.}(2024)\citenamefont
  {Pankovich}, \citenamefont {Kan}, \citenamefont {Wan}, \citenamefont
  {Ostmann}, \citenamefont {Neville}, \citenamefont {Omkar}, \citenamefont
  {Sohbi},\ and\ \citenamefont {Br{\'a}dler}}]{Pankovich_Phys.Rev.Lett._2024}%
  \BibitemOpen
  \bibfield  {author} {\bibinfo {author} {\bibfnamefont {B.}~\bibnamefont
  {Pankovich}}, \bibinfo {author} {\bibfnamefont {A.}~\bibnamefont {Kan}},
  \bibinfo {author} {\bibfnamefont {K.~H.}\ \bibnamefont {Wan}}, \bibinfo
  {author} {\bibfnamefont {M.}~\bibnamefont {Ostmann}}, \bibinfo {author}
  {\bibfnamefont {A.}~\bibnamefont {Neville}}, \bibinfo {author} {\bibfnamefont
  {S.}~\bibnamefont {Omkar}}, \bibinfo {author} {\bibfnamefont
  {A.}~\bibnamefont {Sohbi}},\ and\ \bibinfo {author} {\bibfnamefont
  {K.}~\bibnamefont {Br{\'a}dler}},\ }\bibfield  {title} {\bibinfo {title}
  {High-{{Photon-Loss Threshold Quantum Computing Using GHZ-State
  Measurements}}},\ }\href {https://doi.org/10.1103/PhysRevLett.133.050604}
  {\bibfield  {journal} {\bibinfo  {journal} {Physical Review Letters}\
  }\textbf {\bibinfo {volume} {133}},\ \bibinfo {pages} {050604} (\bibinfo
  {year} {2024})}\BibitemShut {NoStop}%
\bibitem [{\citenamefont {Varnava}\ \emph {et~al.}(2006)\citenamefont
  {Varnava}, \citenamefont {Browne},\ and\ \citenamefont
  {Rudolph}}]{Varnava_Phys.Rev.Lett.2006}%
  \BibitemOpen
  \bibfield  {author} {\bibinfo {author} {\bibfnamefont {M.}~\bibnamefont
  {Varnava}}, \bibinfo {author} {\bibfnamefont {D.~E.}\ \bibnamefont
  {Browne}},\ and\ \bibinfo {author} {\bibfnamefont {T.}~\bibnamefont
  {Rudolph}},\ }\bibfield  {title} {\bibinfo {title} {Loss {{Tolerance}} in
  {{One-Way Quantum Computation}} via {{Counterfactual Error Correction}}},\
  }\href {https://doi.org/10.1103/PhysRevLett.97.120501} {\bibfield  {journal}
  {\bibinfo  {journal} {Physical Review Letters}\ }\textbf {\bibinfo {volume}
  {97}},\ \bibinfo {pages} {120501} (\bibinfo {year} {2006})}\BibitemShut
  {NoStop}%
\bibitem [{\citenamefont {Varnava}\ \emph {et~al.}(2007)\citenamefont
  {Varnava}, \citenamefont {Browne},\ and\ \citenamefont
  {Rudolph}}]{Varnava_NewJ.Phys.2007}%
  \BibitemOpen
  \bibfield  {author} {\bibinfo {author} {\bibfnamefont {M.}~\bibnamefont
  {Varnava}}, \bibinfo {author} {\bibfnamefont {D.~E.}\ \bibnamefont
  {Browne}},\ and\ \bibinfo {author} {\bibfnamefont {T.}~\bibnamefont
  {Rudolph}},\ }\bibfield  {title} {\bibinfo {title} {Loss tolerant linear
  optical quantum memory by measurement-based quantum computing},\ }\href
  {https://doi.org/10.1088/1367-2630/9/6/203} {\bibfield  {journal} {\bibinfo
  {journal} {New Journal of Physics}\ }\textbf {\bibinfo {volume} {9}},\
  \bibinfo {pages} {203} (\bibinfo {year} {2007})}\BibitemShut {NoStop}%
\bibitem [{\citenamefont {Azuma}\ \emph {et~al.}(2015)\citenamefont {Azuma},
  \citenamefont {Tamaki},\ and\ \citenamefont {Lo}}]{Azuma_Nat.Commun.2015}%
  \BibitemOpen
  \bibfield  {author} {\bibinfo {author} {\bibfnamefont {K.}~\bibnamefont
  {Azuma}}, \bibinfo {author} {\bibfnamefont {K.}~\bibnamefont {Tamaki}},\ and\
  \bibinfo {author} {\bibfnamefont {H.-K.}\ \bibnamefont {Lo}},\ }\bibfield
  {title} {\bibinfo {title} {All-photonic quantum repeaters},\ }\href
  {https://doi.org/10.1038/ncomms7787} {\bibfield  {journal} {\bibinfo
  {journal} {Nature Communications}\ }\textbf {\bibinfo {volume} {6}},\
  \bibinfo {pages} {6787} (\bibinfo {year} {2015})}\BibitemShut {NoStop}%
\bibitem [{\citenamefont {Hilaire}\ \emph
  {et~al.}(2023{\natexlab{a}})\citenamefont {Hilaire}, \citenamefont {Castor},
  \citenamefont {Barnes}, \citenamefont {Economou},\ and\ \citenamefont
  {Grosshans}}]{Hilaire_PRXQuantum_2023}%
  \BibitemOpen
  \bibfield  {author} {\bibinfo {author} {\bibfnamefont {P.}~\bibnamefont
  {Hilaire}}, \bibinfo {author} {\bibfnamefont {Y.}~\bibnamefont {Castor}},
  \bibinfo {author} {\bibfnamefont {E.}~\bibnamefont {Barnes}}, \bibinfo
  {author} {\bibfnamefont {S.~E.}\ \bibnamefont {Economou}},\ and\ \bibinfo
  {author} {\bibfnamefont {F.}~\bibnamefont {Grosshans}},\ }\bibfield  {title}
  {\bibinfo {title} {Linear {{Optical Logical Bell State Measurements}} with
  {{Optimal Loss-Tolerance Threshold}}},\ }\href
  {https://doi.org/10.1103/PRXQuantum.4.040322} {\bibfield  {journal} {\bibinfo
   {journal} {PRX Quantum}\ }\textbf {\bibinfo {volume} {4}},\ \bibinfo {pages}
  {040322} (\bibinfo {year} {2023}{\natexlab{a}})}\BibitemShut {NoStop}%
\bibitem [{\citenamefont {Bartolucci}\ \emph {et~al.}(2025)\citenamefont
  {Bartolucci}, \citenamefont {Bell}, \citenamefont {Bombin}, \citenamefont
  {Birchall}, \citenamefont {Bulmer}, \citenamefont {Dawson}, \citenamefont
  {Farrelly}, \citenamefont {Gartenstein}, \citenamefont {{Gimeno-Segovia}},
  \citenamefont {Litinski}, \citenamefont {Liu}, \citenamefont {Knegjens},
  \citenamefont {Nickerson}, \citenamefont {Olivo}, \citenamefont {Pant},
  \citenamefont {Patil}, \citenamefont {Roberts}, \citenamefont {Rudolph},
  \citenamefont {Sparrow}, \citenamefont {Tuckett},\ and\ \citenamefont
  {Veitia}}]{Bartolucci_2025}%
  \BibitemOpen
  \bibfield  {author} {\bibinfo {author} {\bibfnamefont {S.}~\bibnamefont
  {Bartolucci}}, \bibinfo {author} {\bibfnamefont {T.}~\bibnamefont {Bell}},
  \bibinfo {author} {\bibfnamefont {H.}~\bibnamefont {Bombin}}, \bibinfo
  {author} {\bibfnamefont {P.}~\bibnamefont {Birchall}}, \bibinfo {author}
  {\bibfnamefont {J.}~\bibnamefont {Bulmer}}, \bibinfo {author} {\bibfnamefont
  {C.}~\bibnamefont {Dawson}}, \bibinfo {author} {\bibfnamefont
  {T.}~\bibnamefont {Farrelly}}, \bibinfo {author} {\bibfnamefont
  {S.}~\bibnamefont {Gartenstein}}, \bibinfo {author} {\bibfnamefont
  {M.}~\bibnamefont {{Gimeno-Segovia}}}, \bibinfo {author} {\bibfnamefont
  {D.}~\bibnamefont {Litinski}}, \bibinfo {author} {\bibfnamefont
  {Y.}~\bibnamefont {Liu}}, \bibinfo {author} {\bibfnamefont {R.}~\bibnamefont
  {Knegjens}}, \bibinfo {author} {\bibfnamefont {N.}~\bibnamefont {Nickerson}},
  \bibinfo {author} {\bibfnamefont {A.}~\bibnamefont {Olivo}}, \bibinfo
  {author} {\bibfnamefont {M.}~\bibnamefont {Pant}}, \bibinfo {author}
  {\bibfnamefont {A.}~\bibnamefont {Patil}}, \bibinfo {author} {\bibfnamefont
  {S.}~\bibnamefont {Roberts}}, \bibinfo {author} {\bibfnamefont
  {T.}~\bibnamefont {Rudolph}}, \bibinfo {author} {\bibfnamefont
  {C.}~\bibnamefont {Sparrow}}, \bibinfo {author} {\bibfnamefont
  {D.}~\bibnamefont {Tuckett}},\ and\ \bibinfo {author} {\bibfnamefont
  {A.}~\bibnamefont {Veitia}},\ }\href
  {https://doi.org/10.48550/arXiv.2506.11975} {\bibinfo {title} {Comparison of
  schemes for highly loss tolerant photonic fusion based quantum computing}}
  (\bibinfo {year} {2025}),\ \Eprint {https://arxiv.org/abs/2506.11975}
  {arXiv:2506.11975 [quant-ph]} \BibitemShut {NoStop}%
\bibitem [{\citenamefont {Yamazaki}\ and\ \citenamefont
  {Azuma}(2025)}]{Yamazaki_Phys.Rev.Lett._2025}%
  \BibitemOpen
  \bibfield  {author} {\bibinfo {author} {\bibfnamefont {T.}~\bibnamefont
  {Yamazaki}}\ and\ \bibinfo {author} {\bibfnamefont {K.}~\bibnamefont
  {Azuma}},\ }\bibfield  {title} {\bibinfo {title} {Linear-{{Optical Fusion
  Boosted}} by {{High-Dimensional Entanglement}}},\ }\href
  {https://doi.org/10.1103/PhysRevLett.134.200801} {\bibfield  {journal}
  {\bibinfo  {journal} {Physical Review Letters}\ }\textbf {\bibinfo {volume}
  {134}},\ \bibinfo {pages} {200801} (\bibinfo {year} {2025})}\BibitemShut
  {NoStop}%
\bibitem [{\citenamefont {Laha}\ and\ \citenamefont {van
  Loock}(2026)}]{Laha__2026}%
  \BibitemOpen
  \bibfield  {author} {\bibinfo {author} {\bibfnamefont {P.}~\bibnamefont
  {Laha}}\ and\ \bibinfo {author} {\bibfnamefont {P.}~\bibnamefont {van
  Loock}},\ }\href {https://doi.org/10.48550/arXiv.2606.29432} {\bibinfo
  {title} {Squeezing-enhanced {{Pairwise Fusion}} of {{Photonic Qudits}}}}
  (\bibinfo {year} {2026}),\ \Eprint {https://arxiv.org/abs/2606.29432}
  {arXiv:2606.29432 [quant-ph]} \BibitemShut {NoStop}%
\bibitem [{\citenamefont {Raussendorf}\ and\ \citenamefont
  {Briegel}(2001)}]{Raussendorf_Phys.Rev.Lett._2001}%
  \BibitemOpen
  \bibfield  {author} {\bibinfo {author} {\bibfnamefont {R.}~\bibnamefont
  {Raussendorf}}\ and\ \bibinfo {author} {\bibfnamefont {H.~J.}\ \bibnamefont
  {Briegel}},\ }\bibfield  {title} {\bibinfo {title} {A {{One-Way Quantum
  Computer}}},\ }\href {https://doi.org/10.1103/PhysRevLett.86.5188} {\bibfield
   {journal} {\bibinfo  {journal} {Physical Review Letters}\ }\textbf {\bibinfo
  {volume} {86}},\ \bibinfo {pages} {5188} (\bibinfo {year}
  {2001})}\BibitemShut {NoStop}%
\bibitem [{\citenamefont {Briegel}\ \emph {et~al.}(2009)\citenamefont
  {Briegel}, \citenamefont {Browne}, \citenamefont {D{\"u}r}, \citenamefont
  {Raussendorf},\ and\ \citenamefont {{Van den
  Nest}}}]{Briegel_NaturePhys_2009}%
  \BibitemOpen
  \bibfield  {author} {\bibinfo {author} {\bibfnamefont {H.~J.}\ \bibnamefont
  {Briegel}}, \bibinfo {author} {\bibfnamefont {D.~E.}\ \bibnamefont {Browne}},
  \bibinfo {author} {\bibfnamefont {W.}~\bibnamefont {D{\"u}r}}, \bibinfo
  {author} {\bibfnamefont {R.}~\bibnamefont {Raussendorf}},\ and\ \bibinfo
  {author} {\bibfnamefont {M.}~\bibnamefont {{Van den Nest}}},\ }\bibfield
  {title} {\bibinfo {title} {Measurement-based quantum computation},\ }\href
  {https://doi.org/10.1038/nphys1157} {\bibfield  {journal} {\bibinfo
  {journal} {Nature Physics}\ }\textbf {\bibinfo {volume} {5}},\ \bibinfo
  {pages} {19} (\bibinfo {year} {2009})}\BibitemShut {NoStop}%
\bibitem [{\citenamefont {Hilaire}\ \emph
  {et~al.}(2023{\natexlab{b}})\citenamefont {Hilaire}, \citenamefont {Vidro},
  \citenamefont {Eisenberg},\ and\ \citenamefont
  {Economou}}]{Hilaire_Quantum2023}%
  \BibitemOpen
  \bibfield  {author} {\bibinfo {author} {\bibfnamefont {P.}~\bibnamefont
  {Hilaire}}, \bibinfo {author} {\bibfnamefont {L.}~\bibnamefont {Vidro}},
  \bibinfo {author} {\bibfnamefont {H.~S.}\ \bibnamefont {Eisenberg}},\ and\
  \bibinfo {author} {\bibfnamefont {S.~E.}\ \bibnamefont {Economou}},\
  }\bibfield  {title} {\bibinfo {title} {Near-deterministic hybrid generation
  of arbitrary photonic graph states using a single quantum emitter and linear
  optics},\ }\href {https://doi.org/10.22331/q-2023-04-27-992} {\bibfield
  {journal} {\bibinfo  {journal} {Quantum}\ }\textbf {\bibinfo {volume} {7}},\
  \bibinfo {pages} {992} (\bibinfo {year} {2023}{\natexlab{b}})},\ \Eprint
  {https://arxiv.org/abs/2205.09750} {arXiv:2205.09750 [quant-ph]} \BibitemShut
  {NoStop}%
\bibitem [{\citenamefont {Paesani}\ \emph {et~al.}(2021)\citenamefont
  {Paesani}, \citenamefont {Bulmer}, \citenamefont {Jones}, \citenamefont
  {Santagati},\ and\ \citenamefont {Laing}}]{Paesani_Phys.Rev.Lett.2021}%
  \BibitemOpen
  \bibfield  {author} {\bibinfo {author} {\bibfnamefont {S.}~\bibnamefont
  {Paesani}}, \bibinfo {author} {\bibfnamefont {J.~F.~F.}\ \bibnamefont
  {Bulmer}}, \bibinfo {author} {\bibfnamefont {A.~E.}\ \bibnamefont {Jones}},
  \bibinfo {author} {\bibfnamefont {R.}~\bibnamefont {Santagati}},\ and\
  \bibinfo {author} {\bibfnamefont {A.}~\bibnamefont {Laing}},\ }\bibfield
  {title} {\bibinfo {title} {Scheme for {{Universal High-Dimensional Quantum
  Computation}} with {{Linear Optics}}},\ }\href
  {https://doi.org/10.1103/PhysRevLett.126.230504} {\bibfield  {journal}
  {\bibinfo  {journal} {Physical Review Letters}\ }\textbf {\bibinfo {volume}
  {126}},\ \bibinfo {pages} {230504} (\bibinfo {year} {2021})}\BibitemShut
  {NoStop}%
\bibitem [{\citenamefont {Chin}\ \emph {et~al.}(2024)\citenamefont {Chin},
  \citenamefont {Ryu},\ and\ \citenamefont {Kim}}]{Chin_Phys.Rev.Lett._2024}%
  \BibitemOpen
  \bibfield  {author} {\bibinfo {author} {\bibfnamefont {S.}~\bibnamefont
  {Chin}}, \bibinfo {author} {\bibfnamefont {J.}~\bibnamefont {Ryu}},\ and\
  \bibinfo {author} {\bibfnamefont {Y.-S.}\ \bibnamefont {Kim}},\ }\bibfield
  {title} {\bibinfo {title} {Exponentially {{Enhanced Scheme}} for the
  {{Heralded Qudit Greenberger-Horne-Zeilinger State}} in {{Linear Optics}}},\
  }\href {https://doi.org/10.1103/PhysRevLett.133.253601} {\bibfield  {journal}
  {\bibinfo  {journal} {Physical Review Letters}\ }\textbf {\bibinfo {volume}
  {133}},\ \bibinfo {pages} {253601} (\bibinfo {year} {2024})}\BibitemShut
  {NoStop}%
\bibitem [{\citenamefont {Luo}\ \emph {et~al.}(2019)\citenamefont {Luo},
  \citenamefont {Zhong}, \citenamefont {Erhard}, \citenamefont {Wang},
  \citenamefont {Peng}, \citenamefont {Krenn}, \citenamefont {Jiang},
  \citenamefont {Li}, \citenamefont {Liu}, \citenamefont {Lu}, \citenamefont
  {Zeilinger},\ and\ \citenamefont {Pan}}]{Luo_Phys.Rev.Lett._2019}%
  \BibitemOpen
  \bibfield  {author} {\bibinfo {author} {\bibfnamefont {Y.-H.}\ \bibnamefont
  {Luo}}, \bibinfo {author} {\bibfnamefont {H.-S.}\ \bibnamefont {Zhong}},
  \bibinfo {author} {\bibfnamefont {M.}~\bibnamefont {Erhard}}, \bibinfo
  {author} {\bibfnamefont {X.-L.}\ \bibnamefont {Wang}}, \bibinfo {author}
  {\bibfnamefont {L.-C.}\ \bibnamefont {Peng}}, \bibinfo {author}
  {\bibfnamefont {M.}~\bibnamefont {Krenn}}, \bibinfo {author} {\bibfnamefont
  {X.}~\bibnamefont {Jiang}}, \bibinfo {author} {\bibfnamefont
  {L.}~\bibnamefont {Li}}, \bibinfo {author} {\bibfnamefont {N.-L.}\
  \bibnamefont {Liu}}, \bibinfo {author} {\bibfnamefont {C.-Y.}\ \bibnamefont
  {Lu}}, \bibinfo {author} {\bibfnamefont {A.}~\bibnamefont {Zeilinger}},\ and\
  \bibinfo {author} {\bibfnamefont {J.-W.}\ \bibnamefont {Pan}},\ }\bibfield
  {title} {\bibinfo {title} {Quantum {{Teleportation}} in {{High
  Dimensions}}},\ }\href {https://doi.org/10.1103/PhysRevLett.123.070505}
  {\bibfield  {journal} {\bibinfo  {journal} {Physical Review Letters}\
  }\textbf {\bibinfo {volume} {123}},\ \bibinfo {pages} {070505} (\bibinfo
  {year} {2019})}\BibitemShut {NoStop}%
\bibitem [{\citenamefont {Zhang}\ \emph {et~al.}(2019)\citenamefont {Zhang},
  \citenamefont {Chen}, \citenamefont {Cui}, \citenamefont {Dowling},
  \citenamefont {Ou},\ and\ \citenamefont {Byrnes}}]{Zhang_Phys.Rev.A_2019}%
  \BibitemOpen
  \bibfield  {author} {\bibinfo {author} {\bibfnamefont {C.}~\bibnamefont
  {Zhang}}, \bibinfo {author} {\bibfnamefont {J.~F.}\ \bibnamefont {Chen}},
  \bibinfo {author} {\bibfnamefont {C.}~\bibnamefont {Cui}}, \bibinfo {author}
  {\bibfnamefont {J.~P.}\ \bibnamefont {Dowling}}, \bibinfo {author}
  {\bibfnamefont {Z.~Y.}\ \bibnamefont {Ou}},\ and\ \bibinfo {author}
  {\bibfnamefont {T.}~\bibnamefont {Byrnes}},\ }\bibfield  {title} {\bibinfo
  {title} {Quantum teleportation of photonic qudits using linear optics},\
  }\href {https://doi.org/10.1103/PhysRevA.100.032330} {\bibfield  {journal}
  {\bibinfo  {journal} {Physical Review A}\ }\textbf {\bibinfo {volume}
  {100}},\ \bibinfo {pages} {032330} (\bibinfo {year} {2019})}\BibitemShut
  {NoStop}%
\bibitem [{\citenamefont {Hu}\ \emph {et~al.}(2020)\citenamefont {Hu},
  \citenamefont {Zhang}, \citenamefont {Liu}, \citenamefont {Cai},
  \citenamefont {Ye}, \citenamefont {Guo}, \citenamefont {Xing}, \citenamefont
  {Huang}, \citenamefont {Huang}, \citenamefont {Li},\ and\ \citenamefont
  {Guo}}]{Hu_Phys.Rev.Lett._2020}%
  \BibitemOpen
  \bibfield  {author} {\bibinfo {author} {\bibfnamefont {X.-M.}\ \bibnamefont
  {Hu}}, \bibinfo {author} {\bibfnamefont {C.}~\bibnamefont {Zhang}}, \bibinfo
  {author} {\bibfnamefont {B.-H.}\ \bibnamefont {Liu}}, \bibinfo {author}
  {\bibfnamefont {Y.}~\bibnamefont {Cai}}, \bibinfo {author} {\bibfnamefont
  {X.-J.}\ \bibnamefont {Ye}}, \bibinfo {author} {\bibfnamefont
  {Y.}~\bibnamefont {Guo}}, \bibinfo {author} {\bibfnamefont {W.-B.}\
  \bibnamefont {Xing}}, \bibinfo {author} {\bibfnamefont {C.-X.}\ \bibnamefont
  {Huang}}, \bibinfo {author} {\bibfnamefont {Y.-F.}\ \bibnamefont {Huang}},
  \bibinfo {author} {\bibfnamefont {C.-F.}\ \bibnamefont {Li}},\ and\ \bibinfo
  {author} {\bibfnamefont {G.-C.}\ \bibnamefont {Guo}},\ }\bibfield  {title}
  {\bibinfo {title} {Experimental {{High-Dimensional Quantum Teleportation}}},\
  }\href {https://doi.org/10.1103/PhysRevLett.125.230501} {\bibfield  {journal}
  {\bibinfo  {journal} {Physical Review Letters}\ }\textbf {\bibinfo {volume}
  {125}},\ \bibinfo {pages} {230501} (\bibinfo {year} {2020})}\BibitemShut
  {NoStop}%
\bibitem [{\citenamefont {Bharos}\ \emph {et~al.}(2025)\citenamefont {Bharos},
  \citenamefont {Markovich},\ and\ \citenamefont
  {Borregaard}}]{Bharos_Quantum_2025}%
  \BibitemOpen
  \bibfield  {author} {\bibinfo {author} {\bibfnamefont {N.}~\bibnamefont
  {Bharos}}, \bibinfo {author} {\bibfnamefont {L.}~\bibnamefont {Markovich}},\
  and\ \bibinfo {author} {\bibfnamefont {J.}~\bibnamefont {Borregaard}},\
  }\bibfield  {title} {\bibinfo {title} {Efficient {{High-Dimensional Entangled
  State Analyzer}} with {{Linear Optics}}},\ }\href
  {https://doi.org/10.22331/q-2025-04-18-1711} {\bibfield  {journal} {\bibinfo
  {journal} {Quantum}\ }\textbf {\bibinfo {volume} {9}},\ \bibinfo {pages}
  {1711} (\bibinfo {year} {2025})}\BibitemShut {NoStop}%
\bibitem [{\citenamefont {{\"U}st{\"u}n}\ \emph {et~al.}(2025)\citenamefont
  {{\"U}st{\"u}n}, \citenamefont {Rieffel}, \citenamefont {Devitt},\ and\
  \citenamefont {Saied}}]{Ustun_Phys.Rev.Appl._2025}%
  \BibitemOpen
  \bibfield  {author} {\bibinfo {author} {\bibfnamefont {G.}~\bibnamefont
  {{\"U}st{\"u}n}}, \bibinfo {author} {\bibfnamefont {E.~G.}\ \bibnamefont
  {Rieffel}}, \bibinfo {author} {\bibfnamefont {S.~J.}\ \bibnamefont
  {Devitt}},\ and\ \bibinfo {author} {\bibfnamefont {J.}~\bibnamefont
  {Saied}},\ }\bibfield  {title} {\bibinfo {title} {Fusion for high-dimensional
  linear-optical quantum computing with improved success probability},\ }\href
  {https://doi.org/10.1103/l7bg-hc8c} {\bibfield  {journal} {\bibinfo
  {journal} {Physical Review Applied}\ }\textbf {\bibinfo {volume} {24}},\
  \bibinfo {pages} {044024} (\bibinfo {year} {2025})}\BibitemShut {NoStop}%
\bibitem [{\citenamefont {Raissi}\ \emph {et~al.}(2024)\citenamefont {Raissi},
  \citenamefont {Barnes},\ and\ \citenamefont
  {Economou}}]{Raissi_PRXQuantum2024}%
  \BibitemOpen
  \bibfield  {author} {\bibinfo {author} {\bibfnamefont {Z.}~\bibnamefont
  {Raissi}}, \bibinfo {author} {\bibfnamefont {E.}~\bibnamefont {Barnes}},\
  and\ \bibinfo {author} {\bibfnamefont {S.~E.}\ \bibnamefont {Economou}},\
  }\bibfield  {title} {\bibinfo {title} {Deterministic {{Generation}} of
  {{Qudit Photonic Graph States}} from {{Quantum Emitters}}},\ }\href
  {https://doi.org/10.1103/PRXQuantum.5.020346} {\bibfield  {journal} {\bibinfo
   {journal} {PRX Quantum}\ }\textbf {\bibinfo {volume} {5}},\ \bibinfo {pages}
  {020346} (\bibinfo {year} {2024})}\BibitemShut {NoStop}%
\bibitem [{\citenamefont {Paesani}\ and\ \citenamefont
  {Brown}(2023)}]{Paesani_Phys.Rev.Lett._2023}%
  \BibitemOpen
  \bibfield  {author} {\bibinfo {author} {\bibfnamefont {S.}~\bibnamefont
  {Paesani}}\ and\ \bibinfo {author} {\bibfnamefont {B.~J.}\ \bibnamefont
  {Brown}},\ }\bibfield  {title} {\bibinfo {title} {High-{{Threshold Quantum
  Computing}} by {{Fusing One-Dimensional Cluster States}}},\ }\href
  {https://doi.org/10.1103/PhysRevLett.131.120603} {\bibfield  {journal}
  {\bibinfo  {journal} {Physical Review Letters}\ }\textbf {\bibinfo {volume}
  {131}},\ \bibinfo {pages} {120603} (\bibinfo {year} {2023})}\BibitemShut
  {NoStop}%
\bibitem [{\citenamefont {L{\"o}bl}\ \emph
  {et~al.}(2024{\natexlab{a}})\citenamefont {L{\"o}bl}, \citenamefont
  {Paesani},\ and\ \citenamefont {S{\o}rensen}}]{Lobl_Quantum_2024}%
  \BibitemOpen
  \bibfield  {author} {\bibinfo {author} {\bibfnamefont {M.~C.}\ \bibnamefont
  {L{\"o}bl}}, \bibinfo {author} {\bibfnamefont {S.}~\bibnamefont {Paesani}},\
  and\ \bibinfo {author} {\bibfnamefont {A.~S.}\ \bibnamefont {S{\o}rensen}},\
  }\bibfield  {title} {\bibinfo {title} {Loss-tolerant architecture for quantum
  computing with quantum emitters},\ }\href
  {https://doi.org/10.22331/q-2024-03-28-1302} {\bibfield  {journal} {\bibinfo
  {journal} {Quantum}\ }\textbf {\bibinfo {volume} {8}},\ \bibinfo {pages}
  {1302} (\bibinfo {year} {2024}{\natexlab{a}})},\ \Eprint
  {https://arxiv.org/abs/2304.03796} {arXiv:2304.03796 [quant-ph]} \BibitemShut
  {NoStop}%
\bibitem [{\citenamefont {Wein}\ \emph {et~al.}(2025)\citenamefont {Wein},
  \citenamefont {{Goubault de Brugi{\`e}re}}, \citenamefont {Music},
  \citenamefont {Senellart}, \citenamefont {Bourdoncle},\ and\ \citenamefont
  {Mansfield}}]{Wein_PRXQuantum_2025}%
  \BibitemOpen
  \bibfield  {author} {\bibinfo {author} {\bibfnamefont {S.~C.}\ \bibnamefont
  {Wein}}, \bibinfo {author} {\bibfnamefont {T.}~\bibnamefont {{Goubault de
  Brugi{\`e}re}}}, \bibinfo {author} {\bibfnamefont {L.}~\bibnamefont {Music}},
  \bibinfo {author} {\bibfnamefont {P.}~\bibnamefont {Senellart}}, \bibinfo
  {author} {\bibfnamefont {B.}~\bibnamefont {Bourdoncle}},\ and\ \bibinfo
  {author} {\bibfnamefont {S.}~\bibnamefont {Mansfield}},\ }\bibfield  {title}
  {\bibinfo {title} {Minimizing {{Resource Overhead}} in {{Fusion-Based Quantum
  Computation Using Hybrid Spin-Photon Devices}}},\ }\href
  {https://doi.org/10.1103/8l5g-x3b7} {\bibfield  {journal} {\bibinfo
  {journal} {PRX Quantum}\ }\textbf {\bibinfo {volume} {6}},\ \bibinfo {pages}
  {040362} (\bibinfo {year} {2025})}\BibitemShut {NoStop}%
\bibitem [{\citenamefont {Chan}\ \emph {et~al.}(2026)\citenamefont {Chan},
  \citenamefont {Capatos}, \citenamefont {Lodahl}, \citenamefont
  {S{\o}rensen},\ and\ \citenamefont {Paesani}}]{Chan_npjQuantumInf_2026}%
  \BibitemOpen
  \bibfield  {author} {\bibinfo {author} {\bibfnamefont {M.~L.}\ \bibnamefont
  {Chan}}, \bibinfo {author} {\bibfnamefont {A.~A.}\ \bibnamefont {Capatos}},
  \bibinfo {author} {\bibfnamefont {P.}~\bibnamefont {Lodahl}}, \bibinfo
  {author} {\bibfnamefont {A.~S.}\ \bibnamefont {S{\o}rensen}},\ and\ \bibinfo
  {author} {\bibfnamefont {S.}~\bibnamefont {Paesani}},\ }\bibfield  {title}
  {\bibinfo {title} {Practical blueprint for low-depth photonic quantum
  computing with quantum dots},\ }\href
  {https://doi.org/10.1038/s41534-026-01258-3} {\bibfield  {journal} {\bibinfo
  {journal} {npj Quantum Information}\ }\textbf {\bibinfo {volume} {12}},\
  \bibinfo {pages} {155} (\bibinfo {year} {2026})}\BibitemShut {NoStop}%
\bibitem [{Note1()}]{Note1}%
  \BibitemOpen
  \bibinfo {note} {Compared with \cite {Yamazaki_Phys.Rev.Lett._2025}, we
  replace $\ket {i+j}$ with $\ket {-i-j}$ in the definition of $\ket {C}$. This
  makes the state symmetric under permutations of the three qudits and allows
  it to be represented by a single 3-leg qudit X spider.}\BibitemShut {Stop}%
\bibitem [{\citenamefont {Coecke}\ and\ \citenamefont
  {Duncan}(2011)}]{Coecke_NewJ.Phys._2011}%
  \BibitemOpen
  \bibfield  {author} {\bibinfo {author} {\bibfnamefont {B.}~\bibnamefont
  {Coecke}}\ and\ \bibinfo {author} {\bibfnamefont {R.}~\bibnamefont
  {Duncan}},\ }\bibfield  {title} {\bibinfo {title} {Interacting quantum
  observables: Categorical algebra and diagrammatics},\ }\href
  {https://doi.org/10.1088/1367-2630/13/4/043016} {\bibfield  {journal}
  {\bibinfo  {journal} {New Journal of Physics}\ }\textbf {\bibinfo {volume}
  {13}},\ \bibinfo {pages} {043016} (\bibinfo {year} {2011})}\BibitemShut
  {NoStop}%
\bibitem [{\citenamefont {{van de Wetering}}(2020)}]{vandeWetering_2020}%
  \BibitemOpen
  \bibfield  {author} {\bibinfo {author} {\bibfnamefont {J.}~\bibnamefont {{van
  de Wetering}}},\ }\href@noop {} {\bibinfo {title} {{{ZX}}-calculus for the
  working quantum computer scientist}} (\bibinfo {year} {2020}),\ \Eprint
  {https://arxiv.org/abs/2012.13966} {arXiv:2012.13966 [quant-ph]} \BibitemShut
  {NoStop}%
\bibitem [{Note2()}]{Note2}%
  \BibitemOpen
  \bibinfo {note} {We define qudit X spiders using $\bra {\protect \tilde
  {-j}}$ instead of $\bra {\protect \tilde {j}}$ similar to the notation
  in~\cite {deBeaudrap_49thInt.Symp.Math.Found.Comput.Sci.MFCS2024_2024} to
  make diagrams symmetric under interchanges of kets and bras. In addition, we
  let a phase of a qudit spider be a scalar, instead of a vector.}\BibitemShut
  {Stop}%
\bibitem [{\citenamefont {Bombin}\ \emph {et~al.}(2024)\citenamefont {Bombin},
  \citenamefont {Litinski}, \citenamefont {Nickerson}, \citenamefont
  {Pastawski},\ and\ \citenamefont {Roberts}}]{Bombin_Quantum2024}%
  \BibitemOpen
  \bibfield  {author} {\bibinfo {author} {\bibfnamefont {H.}~\bibnamefont
  {Bombin}}, \bibinfo {author} {\bibfnamefont {D.}~\bibnamefont {Litinski}},
  \bibinfo {author} {\bibfnamefont {N.}~\bibnamefont {Nickerson}}, \bibinfo
  {author} {\bibfnamefont {F.}~\bibnamefont {Pastawski}},\ and\ \bibinfo
  {author} {\bibfnamefont {S.}~\bibnamefont {Roberts}},\ }\bibfield  {title}
  {\bibinfo {title} {Unifying flavors of fault tolerance with the {{ZX}}
  calculus},\ }\href {https://doi.org/10.22331/q-2024-06-18-1379} {\bibfield
  {journal} {\bibinfo  {journal} {Quantum}\ }\textbf {\bibinfo {volume} {8}},\
  \bibinfo {pages} {1379} (\bibinfo {year} {2024})}\BibitemShut {NoStop}%
\bibitem [{Note3()}]{Note3}%
  \BibitemOpen
  \bibinfo {note} {In our definition of qudit X spiders, some rewrite rules
  need to be modified as in~\cite
  {deBeaudrap_49thInt.Symp.Math.Found.Comput.Sci.MFCS2024_2024}. For example, a
  phase-free qudit X spider with one input and one output represents $\DOTSB
  \sum@ \slimits@ _{j\in \protect \mathbb {Z}_d} \ket {j}\bra {-j}$, which
  differs from the identity for $d>2$. This would cause little confusion as we
  mainly use rewrite rules for subspace ZX diagrams.}\BibitemShut {Stop}%
\bibitem [{\citenamefont {L{\"o}bl}\ \emph {et~al.}(2025)\citenamefont
  {L{\"o}bl}, \citenamefont {Pettersson}, \citenamefont {Jena}, \citenamefont
  {Dellantonio}, \citenamefont {Paesani},\ and\ \citenamefont
  {S{\o}rensen}}]{Lobl_Phys.Rev.A_2025}%
  \BibitemOpen
  \bibfield  {author} {\bibinfo {author} {\bibfnamefont {M.~C.}\ \bibnamefont
  {L{\"o}bl}}, \bibinfo {author} {\bibfnamefont {L.~A.}\ \bibnamefont
  {Pettersson}}, \bibinfo {author} {\bibfnamefont {A.}~\bibnamefont {Jena}},
  \bibinfo {author} {\bibfnamefont {L.}~\bibnamefont {Dellantonio}}, \bibinfo
  {author} {\bibfnamefont {S.}~\bibnamefont {Paesani}},\ and\ \bibinfo {author}
  {\bibfnamefont {A.~S.}\ \bibnamefont {S{\o}rensen}},\ }\bibfield  {title}
  {\bibinfo {title} {Generating graph states with a single quantum emitter and
  the minimum number of fusions},\ }\href
  {https://doi.org/10.1103/PhysRevA.111.052604} {\bibfield  {journal} {\bibinfo
   {journal} {Physical Review A}\ }\textbf {\bibinfo {volume} {111}},\ \bibinfo
  {pages} {052604} (\bibinfo {year} {2025})}\BibitemShut {NoStop}%
\bibitem [{\citenamefont {Browne}\ and\ \citenamefont
  {Rudolph}(2005)}]{Browne_Phys.Rev.Lett.2005}%
  \BibitemOpen
  \bibfield  {author} {\bibinfo {author} {\bibfnamefont {D.~E.}\ \bibnamefont
  {Browne}}\ and\ \bibinfo {author} {\bibfnamefont {T.}~\bibnamefont
  {Rudolph}},\ }\bibfield  {title} {\bibinfo {title} {Resource-{{Efficient
  Linear Optical Quantum Computation}}},\ }\href
  {https://doi.org/10.1103/PhysRevLett.95.010501} {\bibfield  {journal}
  {\bibinfo  {journal} {Physical Review Letters}\ }\textbf {\bibinfo {volume}
  {95}},\ \bibinfo {pages} {010501} (\bibinfo {year} {2005})}\BibitemShut
  {NoStop}%
\bibitem [{\citenamefont {Duan}\ and\ \citenamefont
  {Raussendorf}(2005)}]{Duan_Phys.Rev.Lett.2005}%
  \BibitemOpen
  \bibfield  {author} {\bibinfo {author} {\bibfnamefont {L.-M.}\ \bibnamefont
  {Duan}}\ and\ \bibinfo {author} {\bibfnamefont {R.}~\bibnamefont
  {Raussendorf}},\ }\bibfield  {title} {\bibinfo {title} {Efficient {{Quantum
  Computation}} with {{Probabilistic Quantum Gates}}},\ }\href
  {https://doi.org/10.1103/PhysRevLett.95.080503} {\bibfield  {journal}
  {\bibinfo  {journal} {Physical Review Letters}\ }\textbf {\bibinfo {volume}
  {95}},\ \bibinfo {pages} {080503} (\bibinfo {year} {2005})}\BibitemShut
  {NoStop}%
\bibitem [{\citenamefont {L{\"o}bl}\ \emph
  {et~al.}(2024{\natexlab{b}})\citenamefont {L{\"o}bl}, \citenamefont
  {Paesani},\ and\ \citenamefont {S{\o}rensen}}]{Lobl_Phys.Rev.Res._2024}%
  \BibitemOpen
  \bibfield  {author} {\bibinfo {author} {\bibfnamefont {M.~C.}\ \bibnamefont
  {L{\"o}bl}}, \bibinfo {author} {\bibfnamefont {S.}~\bibnamefont {Paesani}},\
  and\ \bibinfo {author} {\bibfnamefont {A.~S.}\ \bibnamefont {S{\o}rensen}},\
  }\bibfield  {title} {\bibinfo {title} {Efficient percolation simulations for
  lossy photonic fusion networks},\ }\href
  {https://doi.org/10.1103/PhysRevResearch.6.033273} {\bibfield  {journal}
  {\bibinfo  {journal} {Physical Review Research}\ }\textbf {\bibinfo {volume}
  {6}},\ \bibinfo {pages} {033273} (\bibinfo {year}
  {2024}{\natexlab{b}})}\BibitemShut {NoStop}%
\bibitem [{\citenamefont {{de Beaudrap}}\ and\ \citenamefont
  {East}(2024)}]{deBeaudrap_49thInt.Symp.Math.Found.Comput.Sci.MFCS2024_2024}%
  \BibitemOpen
  \bibfield  {author} {\bibinfo {author} {\bibfnamefont {N.}~\bibnamefont {{de
  Beaudrap}}}\ and\ \bibinfo {author} {\bibfnamefont {R.~D.~P.}\ \bibnamefont
  {East}},\ }\bibfield  {title} {\bibinfo {title} {Simple {{Qudit ZX}} and {{ZH
  Calculi}}, via {{Integrals}}},\ }in\ \href
  {https://doi.org/10.4230/LIPIcs.MFCS.2024.20} {\emph {\bibinfo {booktitle}
  {49th {{International Symposium}} on {{Mathematical Foundations}} of
  {{Computer Science}} ({{MFCS}} 2024)}}},\ \bibinfo {series} {Leibniz
  {{International Proceedings}} in {{Informatics}} ({{LIPIcs}})}, Vol.\
  \bibinfo {volume} {306},\ \bibinfo {editor} {edited by\ \bibinfo {editor}
  {\bibfnamefont {R.}~\bibnamefont {Kr{\'a}lovi{\v c}}}\ and\ \bibinfo {editor}
  {\bibfnamefont {A.}~\bibnamefont {Ku{\v c}era}}}\ (\bibinfo  {publisher}
  {Schloss Dagstuhl -- Leibniz-Zentrum f\"ur Informatik},\ \bibinfo {address}
  {Dagstuhl, Germany},\ \bibinfo {year} {2024})\ pp.\ \bibinfo {pages}
  {20:1--20:20}\BibitemShut {NoStop}%
\end{thebibliography}%

\end{document}